\documentclass[11pt]{article}

\usepackage{epsfig,multicol}
\usepackage{amsmath,amssymb}
\usepackage{bbm}
\usepackage[latin1]{inputenc}
\usepackage{fancyhdr}

\numberwithin{equation}{section}

\newcommand{\nn}{\nonumber\\}
\def\bmm{b^{--}}
\def\bpm{b^{+-}}
\def\bpp{b^{++}}
\def\Chpp{\hat{C}^{++}{}}
\def\Chpm{\hat{C}^{+-}{}}
\def\Chmm{\hat{C}^{--}{}}
\def\Dpmpm{\mathrm{D}^{\pm\pm}}
\def\Do{\mathrm{D}^{0}}
\def\Dmm{\mathrm{D}^{--}}
\def\Dpp{\mathrm{D}^{++}}
\def\epsilonm{\epsilon^-{}}

\def\kpp{\kappa^{++}}

\def\nablapp{\nabla^{++}}
\def\partiall{\overleftarrow{\partial}}

\def\partialr{\overrightarrow{\partial}}
\def\pmm{\partial^{--}}
\def\ppp{\partial^{++}}
\def\Psibm{\bar{\Psi}^-{}}

\def\Psim{\Psi^-{}}

\def\thetabm{\bar{\theta}^-{}}
\def\thetabp{\bar{\theta}^+{}}
\def\thetam{\theta^-{}}
\def\thetap{\theta^+{}}
\def\Vmm{V^{--}}

\def\vmm{v^{--}}
\def\vpp{v^{++}}
\def\vpmm{\varphi^{--}}
\def\iim{\mathrm{i}}
\def\V{V^{++}_{\text{WZ}}}
\def\HR{\mathbbm{H\!\!R}}
\def\ve{\varepsilon}
\def\phib{\bar{\phi}}
\def\Psib{\bar{\Psi}}

\def\thetab{\bar{\theta}}
\def\gb{\bar{g}}
\def\alphad{{\dot{\alpha}}}
\def\betad{{\dot{\beta}}}
\def\gammad{{\dot{\gamma}}}
\def\lambdad{\dot{\lambda}}
\def\Ac{{\cal A}}
\def\Fc{{\cal F}}
\def\Gc{{\cal G}}
\def\Nc{{\cal N}}
\def\Wc{{\cal W}}
\def\L{\Lambda_\epsilon}
\def\l{\lambda_\epsilon}

\title{Non-singlet Q-deformation of the \\
       ${\cal{N}}{=}(1,1)$ gauge multiplet in harmonic superspace}

\author{A. De Castro$\,{}^a$, E. Ivanov$\,{}^b$,
        O. Lechtenfeld$\,{}^a$ and L. Quevedo$\,{}^a$}

\date\empty

\begin{document}

\maketitle
\setcounter{page}{0}

\begin{center}

$^a$ Institut f\"ur Theoretische Physik, Universit\"at Hannover, \\
Appelstrasse 2, 30167 Hannover, Germany. \\
{\tt castro, lechtenf, quevedo@itp.uni-hannover.de}

\vspace{5mm}

$^b$ Bogoliubov Laboratory of Theoretical Physics, JINR, \\
141980 Dubna, Moscow Region, Russia.\\
{\tt eivanov@theor.jinr.ru}
\end{center}

\vspace{10mm}

\thispagestyle{fancy}
\fancyhf{}
\fancyhead[RO]{ITP-UH 08/05 \\ hep-th/0510013}
\renewcommand{\headrulewidth}{0pt}

\begin{abstract}
\noindent{We study a non-anticommutative chiral non-singlet deformation
of the ${\cal N}{=}(1,1)$ abelian gauge multiplet in Euclidean harmonic
superspace with a product ansatz for the deformation matrix,
$C^{(\alpha\beta)}_{(ik)} = c^{(\alpha\beta)}b_{(ik)}$. This choice allows us
to obtain in closed form the gauge transformations and the unbroken
${\cal N}{=}(1,0)$ supersymmetry transformations preserving the
Wess-Zumino gauge, as well as the bosonic sector of the ${\cal
N}{=}(1,0)$ invariant action. This should be contrasted with the generic
choice for which the analogous results are known only to a few orders in
the deformation parameters. As in the case of a singlet deformation,
the bosonic action can be cast in a form where it differs from the free
action merely by a scalar factor. The latter is now given by $\cosh^2
(2\bar\phi\sqrt{c^{\alpha\beta}c_{\alpha\beta}\,b^{ik}b_{ik}}\,)\,$,
with $\bar\phi$ being one of two scalar fields of the ${\cal N}{=}(1,1)$
vector multiplet. We compare our results with previous studies of
non-singlet deformations, including the degenerate case
$b^{(ik)}b_{(ik)}=0$ which preserves the ${\cal N}{=}(1,\frac12)$
fraction of ${\cal N}{=}(1,1)$ supersymmetry.}
\end{abstract}


\newpage

\section{Introduction}
Non-anticommutative ---or nilpotent--- deformations of Euclidean
superspaces naturally emerge in string theory while considering a
low-energy limit of some superstrings in special backgrounds, e.g. a
graviphoton background \cite{deBoer:2003dn, Ooguri:2003qp}.  These
string theory-inspired deformations have recently boosted the interest
in \emph{non-anticommutative} Euclidean supersymmetric field theories,
with properly broken $\Nc=(\frac12, \frac12)$ or $\Nc=(1,1)$
supersymmetries \cite{Seiberg:2003yz} -
\cite{Aldrovandi:2005st}.\footnote{A more complete list of references
can be found e.g.  in \cite{Araki:2005pc}.}  In the superfield approach,
the nilpotent deformations are introduced via Weyl-Moyal product with a
bilinear Poisson operator which is constructed either in terms of the
supercharges, or in terms of the spinor covariant derivatives
\cite{Ferrara1,Klemm,Ferrara:2003xy}. These two options lead to the Q-
or D-deformation, respectively. Q-deformations break supersymmetry but
preserve chirality and, in the $\Nc=(1,1)$ case, also Grassmann harmonic
analyticity. On the other hand, D-deformations are
supersymmetry-preserving, but they break chirality and, generically,
Grassmann harmonic analyticity. Since it is Q-deformations that are
directly implied by string theory (at least for a few well elaborated
cases), it seems tempting to continue studying for a while their various
options and physical consequences thereof, leaving aside the issue of
the precise relation of these options to specific string backgrounds. As
pointed out in \cite{Ivanov:2004hc}, these deformations can e.g. provide
a new mechanisms of soft supersymmetry breaking in some realistic
supersymmetric theories. From the mathematical point of view, they are
also of interest, giving rise to a new variant of q-deformed
supersymmetry \cite{ZupN,Hann}. Their impact on the target space
geometries of supersymmetric sigma models and corresponding scalar
potentials was studied e.g. in \cite{Geom}.

In this paper we shall deal with the $\Nc=(1,1)\rightarrow \Nc=(1,0)$
supersymmetry breaking Q-deformations, the associated Poisson operator
of which reads
\begin{equation}\label{Poisson}
P = -\overleftarrow{Q}^i_\alpha C^{\alpha \beta}_{ik}
\overrightarrow{Q}^k_\beta\,.
\end{equation}
The Moyal product of two superfields is then defined by
\begin{equation}
A\star B=A e^P B\,. \label{MP}
\end{equation}
The deformation parameters $C^{\alpha\beta}_{ij}$ form a constant tensor
which is symmetric under the simultaneous permutation of the Latin and
Greek indices, ${C}^{\alpha\beta}_{ij}={C}^{\beta\alpha}_{ji}$.
Generically, it breaks the full automorphism symmetry $Spin(4)\times$
O(1,1)$\times$ SU(2) $\equiv$ SU(2)$_{\text{L}}\times$
SU(2)$_{\text{R}}\times$ O(1,1)$\times$ SU(2) of the $\Nc=(1,1)$
superalgebra (O(1,1) and SU(2) are the R-symmetry groups) down to
SU(2)$_\text{R}$. The operator \eqref{Poisson} can be split (see for
example \cite{Ferrara:2003xy, Ivanov:2003te}) as follows
\begin{equation}
P = -I \overleftarrow{Q}^i_\alpha \ve^{\alpha \beta}\ve_{ik}
\overrightarrow{Q}^k_\beta -\overleftarrow{Q}^i_\alpha
\hat{C}^{\alpha \beta}_{ik}
\overrightarrow{Q}^k_\beta\,. \label{Generic}
\end{equation}
The first term is $Spin(4)\times$SU(2)-preserving while the second term
involves a SU(2)$_\text{L}\times$SU(2) constant tensor which is
symmetric under the independent permutations of Latin and Greek indices,
$\hat{C}^{\alpha\beta}_{ij}=
\hat{C}^{\beta\alpha}_{ij}=\hat{C}^{\alpha\beta}_{ji}$. For the generic
choice, it fully breaks ``Lorentz'' symmetry SU(2)$_{\text{L}}$ and
R-symmetry SU(2).  Q-deformations induced by the first term only are
called \emph{singlet} or \emph{QS-deformations}, whereas those
associated with the second term can naturally be named
\emph{non-singlet} or \emph{QNS-deformations}.  A key feature of
Q-deformations and D-deformations is the nilpotent nature of the Poisson
operator \eqref{Poisson} which advantageously makes the Moyal product
polynomial. In the case under consideration $P^5=0$ and, as a result,
\begin{equation}
\label{MoyalPoly}
A\star\,B=A\,B+A\,P\,B+\frac12\,A\,P^2\,B
+\frac16\,A\,P^3\,B+\frac1{24}A\,P^4\,B\,.
\end{equation}
A detailed treatment of this Poisson operator in the harmonic superspace
approach was given in \cite{Ivanov:2003te,Ferrara:2004zv}. The Moyal
product associated with it breaks $\Nc=(1,1)$ supersymmetry down to
$\Nc=(1,0)$ ($P$ does not commute with $\bar{Q}_{\dot\alpha\,i}$) but
preserves both the chirality and Grassmann harmonic analyticity of the
involved superfields, as well as the harmonic conditions $D^{\pm\pm}
A=0$, in virtue of the properties
\begin{equation}
\left[D^{\pm}_\alpha , P\right] = 0\,, \qquad
\left[\bar{D}^{\pm}_{\alphad} , P\right] = 0\,, \qquad
\left[D^{\pm\pm} , P\right] = 0
\end{equation}
(see \cite{Ivanov:2003te,Ferrara:2004zv} and Appendix \ref{Appendix1}
for details of the notation).

In the present paper, we study the QNS-deformation of $\Nc=(1,1)$
supersymmetric U(1) vector multiplet in harmonic superspace, thereby
continuing the series of works on QS-deformations of the vector
multiplet \cite{Ferrara:2004zv} and hypermultiplet \cite{Ivanov:2004hc}.
To simplify things, we choose the deformation tensor in the particular
factorizable form
\begin{equation}\label{hatcdecomp}
\hat{C}_{ij}^{\alpha\beta}=b_{(ij)}c^{(\alpha\beta)}\,.
\end{equation}
It breaks the ``Lorentz'' group SU(2)$_{\text{L}}$ and R-symmetry group
SU(2) down to U(1)$_{\text{L}}$ and U(1), respectively. For this choice,
we explicitly present the gauge transformation of the fields in the
Wess-Zumino (WZ) gauge, as well as give a few examples of unbroken
$\Nc=(1,0)$ supersymmetry transformations. We also calculate the bosonic
part of the deformed $\Nc=(1,0)$ supersymmetric action.
The choice \eqref{hatcdecomp} allows us to obtain {\it exact\/} expressions 
for the deformed $\Nc=(1,0)$ gauge and supersymmety transformation laws
and the action, in contrast to the generic choice for which the
corresponding expressions are known as formal series in the deformation 
parameters \cite{Araki:2004mq, Araki:2004cv}. More precisely,
the authors of~\cite{Araki:2004mq, Araki:2004cv} have obtained the all-order
formula for the gauge and susy transformations in power-series form
and explicitly presented the lowest orders.
The unique possibility to deduce all results in closed form
is a major reason why the non-singlet deformation with the ansatz
\eqref{hatcdecomp} is worthy of a detailed study. When turning off some 
components of $\hat{C}_{ij}^{\alpha\beta}$ it is also possible to find exact 
expressions; see for example~\cite{Araki:2005pc} where the exact gauge and 
$\Nc=(1,\frac12)$ supersymmetry transformations are obtained for a special case.

The bosonic action has a structure similar to the QS-deformed action
calculated in \cite{Araki:2004de,Ferrara:2004zv}. Namely, after a
proper field redefinition the Lagrangian proves to differ from the
undeformed one merely by a scalar factor which is a regular function of
the argument
\begin{equation}
X=2\phib\sqrt{b^2 c^2}, \qquad\text{where} \quad
b^2 = b^{ij}b_{ij} \quad\text{and}\quad
c^2 = c^{\alpha\beta}c_{\alpha\beta}\,,
\end{equation}
and $\phib$ is one of two real scalar fields of the gauge $N=(1,1)$
multiplet. The precise relation is as follows
\begin{equation}
S=\int\,d^4x\;  \cosh^2 (2\phib\sqrt{c^2b^2}\,)\left[
-\frac12\varphi\square\phib
+\frac14 d^{ij}d_{ij}
-\frac1{16}\widetilde{F}^{\alpha\beta}\widetilde{F}_{\alpha\beta} \right],
\end{equation}
where $\widetilde{F}_{\alpha\beta} = 2\iim\partial_{(\alpha\dot\alpha}
\widetilde{A}^{\dot\alpha}_{\beta)}$ is a self-dual component of the
gauge field strength, with the potential
$\widetilde{A}_{\dot\alpha\beta}$ possessing the standard abelian gauge
transformation law.\footnote{Recall that at the quadratic level
$\widetilde{F}^{\alpha\beta}\widetilde{F}_{\alpha\beta}$ equals to the
standard gauge field kinetic term $\sim
\widetilde{F}^{\alpha\beta}\widetilde{F}_{\alpha\beta} +
\widetilde{F}^{\dot\alpha\dot\beta}\widetilde{F}_{\dot\alpha\dot\beta}$
modulo a total derivative.} We observe that the only non-trivial
irremovable interaction in the bosonic limit is that between $\bar\phi$
and $\tilde{F}_{\alpha\beta}$, as in the case of QS-deformation.
Expanding our results in powers of $X$ we show that the first-order
terms are comparable with the formal series results obtained in
\cite{Araki:2004mq, Araki:2004cv}.  We
also consider the case of degenerate QNS-deformation preserving the
$\Nc=(1,\frac12)$ portion of supersymmetry \cite{Ivanov:2003te}. It
corresponds to the choice $b^2 =0\,$. We find agreement with a recent
paper \cite{Araki:2005pc} where the same option was treated within the
${\cal N}=1$ superfield formalism.

The basic conventions used throughout the paper are summarized in
Appendix \ref{Appendix1}. Appendix \ref{harmints} contains the list of
useful harmonic integrals.

\section{Generalities of Q-deformed $\Nc=(1,1)$ gauge theory}
\subsection{The structure of generic Q-deformations}
We use the harmonic superspace \cite{Galperin1,Galperin:2001uw}
and work in its left-chiral basis, as
defined in Appendix \ref{Appendix1}, where
\begin{equation}
Q^i_\alpha = \partial^i_\alpha\,, \quad
Q^\pm_\alpha \equiv Q^i_\alpha u^\pm_i
= \pm \partial_{\mp \alpha}\,.
 \end{equation}
In this basis, the Poisson operator \eqref{Poisson}, \eqref{Generic} of
the generic Q-deformation is written as
\begin{equation}
P = -\partiall^i_\alpha C^{\alpha \beta}_{ik}
\partialr^k_\beta\,, \label{YQ}
\end{equation}
where
\begin{equation}
C^{\alpha\beta}_{ik} =
\hat{C}^{(\alpha\beta)}_{(ik)}
+I\,\epsilon^{\alpha\beta}\epsilon_{ik}\,.
\end{equation}
Being expressed in terms of the harmonic projections
\begin{equation}
C^{\pm\pm\, \alpha\beta} =
\hat C^{\pm\pm\,(\alpha\beta)}\,,
\quad C^{\pm\mp\,\alpha\beta} =
\hat C^{\pm\mp\,(\alpha\beta)}
\pm I\,\epsilon^{\alpha\beta}\,,
\quad \hat C^{+-\,(\alpha\beta)} =
\hat C^{-+\,(\alpha\beta)}\,,
\end{equation}
the Poisson operator reads
\begin{equation}
\begin{aligned}
    P=-\partiall_{+\alpha}\Chpp^{\alpha\beta}\partialr_{+\beta}
      -\partiall_{+\alpha}\left( \Chpm^{\alpha\beta}
      +I\varepsilon^{\alpha\beta} \right)\partialr_{-\beta}\\
      -\partiall_{-\alpha}\left( \Chpm^{\alpha\beta}
      -I\varepsilon^{\alpha\beta} \right)\partialr_{+\beta}
      -\partiall_{-\alpha}\Chmm^{\alpha\beta}\partialr_{-\beta}\,.
\end{aligned}
\end{equation}

The Q-deformed commutator of two Grassmann-even and harmonic-analytic
superfields $A$ and $B$ can be calculated using the definition
\eqref{MP}, \eqref{MoyalPoly} of the star product.  For the superfields
commuting with respect to the ordinary product, $\left[A, B \right] =
0$, the deformed commutator is given by
\begin{align}\label{qdefbracket}
\left[\, A,\,B\,\right]_{\star} =&
-2\Bigl[  I\left( \partial^{\alpha}_-A  \partial_{+\alpha}B
- \partial^{\alpha}_+A \partial_{-\alpha}B \right)
+(\partial_{-\alpha}A \partial_{-\beta}B)\Chmm^{\alpha\beta}\cr &
+(\partial_{+\alpha}A \partial_{+\beta}B) \Chpp^{\alpha\beta}
+\, \left(\partial_{-\alpha}A \partial_{+\beta}B
+\partial_{+\alpha}A \partial_{-\beta}B \right)
\Chpm^{\alpha\beta} \Bigl]
\crcr &-3\,[\partial_{-\alpha}(\partial_+)^2 A\,
\partial_{-\beta}(\partial_+)^2 B]
M^{++\,\alpha\beta}\,,
\end{align}
where
\begin{eqnarray}
M^{++\,(\alpha\beta)} &=& \hat C^{+-\,(\alpha \gamma)}
\hat C^{++}_{(\gamma \mu)} \hat C^{+-\,(\mu\beta)}
-\hat C^{++\,(\alpha \gamma)} \hat C^{++}_{(\gamma \mu)}
\hat C^{--\,(\mu\beta)} \nonumber\\
&& -I\left[\hat C^{++\,\alpha}_\gamma
\hat C^{+-\,(\gamma\beta)}
+ \hat C^{++\,\beta}_\gamma \hat C^{+-\,(\gamma\alpha)}\right]
+ I^2\,\hat C^{++\,(\alpha\beta)}\,.\label{Mpp}
\end{eqnarray}
Thus only first- and third-order terms contribute to the star commutator
of the commuting analytic superfields. Since we are interested in the
deformation of Abelian $\Nc=(1,1)$ gauge theory, it will be basically
sufficient to know the relations \eqref{qdefbracket}, \eqref{Mpp}.  Note
that for the special choice
\begin{equation}
\hat C^{(\alpha\beta)}_{(ik)} =
c^{(\alpha\beta)} b_{(ik)}\, \label{Ccb}
\end{equation}
the expression \eqref{Mpp} drastically simplifies to
\begin{equation}
M^{++\,\alpha\beta} =
c^{(\alpha\beta)}\bpp(I^2 - \frac 14 c^2 b^2)\,
\quad\text{with}\quad
c^2 = c^{(\alpha\beta)}c_{(\alpha\beta)}\,,\;\; b^2
= b^{(ik)}b_{(ik)}\,,
\end{equation}
and vanishes for a particular relation between the deformation
parameters, i.e.
\begin{equation}
I^2 = \frac 14 c^2 b^2\,.
\end{equation}

In what follows we shall use just the choice \eqref{Ccb} in the
component calculations since it allows one to obtain all the basic
quantities in a closed form.  While the generic $\hat
C^{\alpha\beta}_{ik}$ fully breaks both SU(2)$_{\text{L}}$ and SU(2)
automorphism symmetries, the particular ansatz \eqref{Ccb} breaks these
symmetries down to U(1)$_{\text{L}}$ and U(1). Thus it is
the maximally symmetric choice for the non-singlet deformation matrix.
A generic matrix $\hat C^{\alpha\beta}_{ik}$ comprises three essential
real parameters as compared to two such parameters in \eqref{Ccb}. In Sect. 4 we shall also
consider a degenerate case of \eqref{Ccb} with $b^2 = 0$.\footnote{The
choice of $b^2=0$ is obviously inconsistent with the standard
SU(2)-covariant reality condition $(b_{ik})^\dagger =
\ve^{ij}\ve^{kl}b_{jl}$ under which $b^2 = 0$ implies $b_{ik} =0\,$. It
is still compatible with a non-vanishing $b_{ik}$ if the reality is
defined with respect to a pseudo-conjugation and the R-symmetry group of
$\Nc=(1,1)$ superalgebra is U(1) from the very beginning
\cite{Ivanov:2003te}. Fortunately, our further consideration does not
depend on whether or not $c^{\alpha\beta}$ and $b^{ik}$ are subject to
any reality condition, which allows us not to care about this issue.}

\subsection{Deformed gauge transformations}
The residual gauge transformations of the component fields of the
Abelian $\Nc=(1,1)$ vector multiplet in the WZ gauge can be found
from the Q-deformed superfield transformation \cite{Ferrara:2004zv}
\begin{equation}
\delta_\Lambda \V = \Dpp\Lambda+[\V, \Lambda]_{\star}\,, \label{DeltaV}
\end{equation}
with $\V$ being the analytic harmonic U(1) superfield gauge connection
and $\Lambda$ the analytic residual gauge parameter satisfying
$D^{+}_\alpha \Lambda = \bar{D}^{+}_{\alphad} \Lambda = 0\,$.

The superfield gauge parameter $\Lambda$ should be chosen so as to
preserve WZ gauge.  In the left-chiral basis, where
$x_A^{\alpha\alphad}= x_L^{\alpha\alphad}
-4\iim\theta^{-\alpha}\thetab^{+\alphad}$ (see Appendix \ref{Appendix1}
for details), $\V$ has the following $\theta$-expansion
\begin{equation}
\V =
v^{++} +\bar\theta^{+}_{\alphad}v^{+\alphad} +(\thetabp)^2 v\, ,
\end{equation}
where
\begin{subequations}
\begin{align}
\label{vpp}&v^{++} = (\theta^+)^2 \phib\, ,\\[3mm]
\label{vp}&v^{+\alphad} = 2\theta^{+\alpha}A_{\alpha}^{\alphad} +
4(\theta^+)^2\Psib^{-\alphad}
-2\iim (\theta^+)^2\theta^{-\alpha}\partial_\alpha^{\alphad}\,\phib\,,\\[3mm]
\label{v}&v = \phi + 4\theta^+\Psi^- + 3(\theta^+)^2 D^{--}
-\iim (\thetap\thetam)\partial^{\alpha\alphad}A_{\alpha\alphad}
+ \theta^{-\alpha}\theta^{+\beta}\,F_{\alpha\beta}\\ &
-(\thetap)^2(\thetam)^2\square\phib+4\iim\,
(\thetap)^2\theta^{-\alpha}\partial_{\alpha\alphad}\Psib^{-\alphad}\,.
\nonumber
\end{align}
\end{subequations}
As the first step in deriving the gauge transformation of the components
of $\V$, we substitute in \eqref{DeltaV} the residual gauge parameter
$\Lambda_0$ used in the undeformed and QS-deformed theories
\cite{Ferrara:2004zv}.  In chiral coordinates it reads
\begin{equation}
\Lambda_0 = \iim a
+2\theta^{-\alpha}\bar\theta^{+\alphad}\partial_{\alpha\alphad}a
-\iim(\thetam)^2(\thetabp)^2 \,\square\,a\,. \label{Lambda0}
\end{equation}
Next, we calculate the star-commutator involving only $\Lambda_0$. Note
that the third-order term in \eqref{qdefbracket} does not contribute to
the gauge transformation of $\V$ in view of the condition
\cite{Ferrara:2004zv}
\begin{equation}
\partial_{+\alpha}\Lambda_0= 0\,.
\end{equation}
Explicit calculations of \eqref{DeltaV} with making use of
\eqref{qdefbracket}, and \eqref{Lambda0} as the gauge parameter gives the
following result
\begin{align}\label{delta0gauge}
&\delta_0\phib=0\,,\nn
&\delta_0\Psib^{-}_{\alphad}=0\,,\nn
&\delta_0 A_{\alpha\alphad}= \partial_{\alpha\alphad}a
+4\,\phib\,\partial_{\beta\alphad} a \hat{C}^{+-\beta}_{\alpha}\,,
\nonumber\\
&\delta_0\phi = 4\partial_{\beta\betad}a\;
A_{\alpha}^{\betad}\hat{C}^{+-\alpha\beta}\,,\nonumber\\
&\delta_0 D^{--}=-\frac{4\iim}3\partial_{\beta\betad}a
\partial_{\alpha}^{\betad}\phib
\hat{C}^{--\alpha\beta}\,,\nonumber\\
&\delta_0\Psi^{-\alpha}=-4\Psib^{-\gammad}\,
\partial_{\beta\dot\gamma}a\, \hat{C}^{+-\alpha\beta}\,.
\end{align}
Here, we omitted the singlet terms ($\sim I$) which can be found in
\cite{Ferrara:2004zv}.  We observe that the transformations
\eqref{delta0gauge}, in contrast to those of the singlet case, do not
preserve WZ gauge because of the appearance of an unwanted dependence on
the harmonic variables $u^{\pm}_i$ in their right-hand sides. This
forces us to choose the gauge parameter $\Lambda$ as a sum of the
`na\"{i}ve' one $\Lambda_0$ and some correction terms
\begin{equation}
    \Lambda=\Lambda_0+\Delta\Lambda\,,
\end{equation}
with
\begin{eqnarray}
&& \Delta\Lambda = \theta^{+}_\alpha \bar\theta^+_{\alphad}\,
\partial^{\alphad}_\beta\,a\,B^{--\alpha\beta}_1
+ (\bar\theta^+)^2 \partial_{\beta\dot\beta}\,a\,
A^{\dot\beta}_\alpha G^{--\alpha\beta} +
(\theta^+)^2(\bar\theta^+)^2 \,\square\,a\, P^{-4} \nn
&& +\, (\bar\theta^+)^2\theta^+_\alpha \,\left[\,\Psib^{-\dot\beta}
\partial_{\beta\dot\beta}\,a\,B_2^{--\alpha\beta} +
\Psib^{+\dot\beta}\partial_{\beta\dot\beta}\,a\,G^{-4\alpha\beta}\,\right]
+(\theta^+)^2(\bar\theta^+)^2\, \partial_{\alpha\alphad}a\,
\partial^{\alphad}_\beta\,\phib\,B^{-4\alpha\beta}_3 \nn
&& +\iim\,\theta^+_\alpha\theta^-_\beta\, (\bar\theta^+)^2\,\square\,a
\,B^{--\alpha\beta}_1 + \iim\, \theta^+_\alpha \theta^-_\gamma
(\bar\theta^+)^2\,\partial_{\beta\dot\lambda}a\,
\partial^{\gamma\lambdad}\phib\,\frac{d}{d\phib}B^{--\alpha\beta}_1\,.
\label{DeltaLambda}
\end{eqnarray}
The coefficients in \eqref{DeltaLambda} are some undetermined functions
of harmonics, the field $\phib$ and deformation parameters. Note that
these coefficients involve both the symmetric and antisymmetric pieces
in the spinor indices. Now we should calculate the correction term to
$\delta_0 \V$,
\begin{equation}
\hat{\delta}\V = \Dpp\Delta \Lambda + [\V, \Delta\Lambda]_\star\,.
\label{hatdeltaV}
\end{equation}
From the structure of $\Delta\Lambda$ we can conclude that only the
lowest order terms in the deformation parameters contribute to the
star-commutator in \eqref{hatdeltaV} calculated according to \eqref{qdefbracket}
and that the term $\sim\hat{C}^{--\alpha\beta}$ is vanishing. Thus we are left with
\begin{align}
[\V, \Delta\Lambda]_\star =&
-2 I\left(\partial^\alpha_-\V\partial_{+\alpha}\Delta\Lambda
-\partial^\alpha_+\V\partial_{-\alpha}\Delta\Lambda\right)
-2\left(\partial_{+\alpha}\V\partial_{+\beta}\Delta\Lambda\right)
\hat{C}^{++\alpha\beta}\\[2mm]\nonumber
&-\,2 \left(\partial_{-\alpha} \V\partial_{+\beta}\Delta\Lambda +
\partial_{+\alpha}\V\partial_{-\beta}
\Delta\Lambda\right)\hat{C}^{+-\alpha\beta}\,.
\end{align}
Now we are ready to find the full gauge transformations of the fields
following from
\begin{equation}\label{deltaV}
    \delta \V =\delta_0 \V + \hat{\delta} \V\,.
\end{equation}
Requiring these full gauge transformations to preserve the WZ gauge
amounts to the following conditions
\begin{subequations}\label{gaugeWZcond}
\begin{align}
\label{gaugeWZconda}\partial^{++}\delta A_{\alpha\alphad} = 0\;&
\leftrightarrow\; \partial^{--}\delta A_{\alpha\alphad}=0\,,\\[3mm]
\label{gaugeWZcondb}\partial^{++}\delta \phi = 0 \; &
\leftrightarrow\; \partial^{--}\delta\phi = 0\,,\\[3mm]
(\partial^{++})^2\delta \Psi^-_\alpha
= 0\;&\leftrightarrow\; \partial^{--}\delta \Psi^-_\alpha = 0\,, \\[3mm]
(\partial^{++})^3\delta D^{--} =
0\;&\leftrightarrow\; \partial^{--}\delta D^{--} = 0\,.
\end{align}
\end{subequations}
After substituting the precise form of the gauge variations, these
conditions fix the unknown harmonic functions in terms of $\phib$,
deformation parameters and harmonics.  After solving them, one can find
the explicit form of the gauge variations.  Unfortunately, it is very
difficult to find the closed solution of these equations for the generic
deformation parameters, though their perturbative solution always exists
as an infinite series in these parameters.

Remarkably, the solution can be found in a closed form if one assumes
the product structure \eqref{Ccb} for the non-singlet part of the
deformation matrix.  In the rest of our paper we will deal just with
this choice, though its possible stringy origin, e.g. as some special
$\Nc=4$ superstring background,\footnote{The stringy interpretation of
the QS-deformation was given in \cite{Ferrara:2004zv}.} still remains to
be revealed.

\section{The precise form of the residual gauge transformations}
In the first two subsections we  present, as instructive examples, the
calculation  of $\delta A_{\alpha\alphad}$ and $\delta \phi$ for the
choice \eqref{Ccb}.  The full list of the residual gauge transformations
is given in the last subsection.

\subsection{Gauge transformation of $A_{\alpha\alphad}$}
The full expression for the variation of $A_{\alpha\alphad}$ following
from \eqref{deltaV} upon using \eqref{DeltaLambda} is
\begin{equation}
\delta A_{\alpha\alphad} = \partial_{\alpha\alphad}a +
4\partial_{\beta\alphad}a \,\phib\, \hat{C}^{+-\beta}_\alpha
+ 2 \partial_{\beta\alphad}a \,
\phib\,B^{--\;\beta}_{1\rho} \hat{C}^{++\;\rho}_\alpha
+\frac12\, \partial_{\beta\alphad}a\,
\partial^{++}B^{--\;\beta}_{1\alpha}\,. \label{211}
\end{equation}
The condition \eqref{gaugeWZconda} amounts to the harmonic equation
\begin{equation}
(\partial^{++})^2B^{--\;\beta}_{1\alpha}
+4\,\phib\,\hat{C}^{++\rho}_\alpha\,\partial^{++}B^{--\;\beta}_{1\rho}
+8\,\phib\,\hat{C}^{++\beta}_\alpha = 0\,,
\end{equation}
which is equivalent to the system
\begin{subequations}
\begin{align}
&\!\!\!\!(\ppp)^2B^{--}
-2\phib \hat{C}^{++\alpha\beta}
\ppp\hat{B}^{--}_{\;\alpha\beta} =0\,,\label{217}\\
&\!\!\!\!(\ppp)^2\hat{B}^{--\alpha\beta}
+ 4\phib\,\hat{C}^{++\alpha\beta}\,\ppp B^{--}+
4\phib\,\hat{C}^{++(\alpha\rho}
\ppp\hat{B}^{--\beta)}_{1\;\rho}
+8\phib\,\hat{C}^{++\alpha\beta}=0\,.
\end{align}
\end{subequations}
Here
\begin{equation}
    {B^{--\;\alpha}_1}_\beta \equiv {\hat{B}^{--\;\alpha}}_{\qquad\beta}
    + \delta^\alpha_{\;\beta}\,B^{--}\,, \quad
    {\hat{B}^{--\;\alpha}}_{\qquad\alpha} = 0\,.
\end{equation}
Though it is easy to obtain a closed equation for
\begin{equation}
\Gc = \ppp B^{--} + 2\,,\label{def1}
\end{equation}
viz.
\begin{equation}
(\ppp)^2 \Gc + 8\phib^2\,(\Chpp)^2\,\Gc = 0\,,
\quad (\Chpp)^2 \equiv
\hat{C}^{++\alpha\beta} \Chpp_{\alpha\beta}\,, \label{219}
\end{equation}
its solution in a close form is very difficult to find without further
simplifications. Equations \eqref{217}, \eqref{219}, for example, can be
solved by iterations to any order in the deformation parameter. Closed
solutions for these and the remaining constraints in \eqref{gaugeWZcond}
can be found in the simplified case \eqref{Ccb}. For this case we can
expand tensor fields over the basis $\{c^{\alpha\beta},
\varepsilon^{\alpha\beta}\}$, for instance
\begin{equation}
B^{--\;\alpha\beta}_1=F^{--}c^{\alpha\beta}
+B^{--}\varepsilon^{\alpha\beta}\,.
\end{equation}
Then, defining
\begin{equation}
\Fc \equiv \ppp F^{--}\,,\label{def2}
\end{equation}
as a consequence of eqs. \eqref{217} we obtain the following equations
\begin{subequations}\label{221}
\begin{eqnarray}
&& \ppp\Gc - 2\,\phib\,c^2\,\bpp\, \Fc = 0\,, \\
&& \ppp\Fc +4\,\phib\,\bpp\,\Gc =0\,,
\end{eqnarray}
\end{subequations}
where $c^2 = c^{\alpha\beta}c_{\alpha\beta}$. Applying  $\partial^{++}$
to \eqref{221} once more, we arrive at the decoupled system
\begin{eqnarray}
&& (a)\;\;(\partial^{++})^2\Gc + (\kpp)^2\,\Gc = 0\,, \nn
&& (b)\;\;(\partial^{++})^2\Fc + (\kpp)^2\,\Fc =0\,,
\end{eqnarray}
where $\kpp=2\,{\phib}\sqrt{2\,c^2\,}\,\bpp$.  These equations are a
sort of the harmonic oscillator ones and they are solved by
\begin{equation}
\Fc = (C_1\, ,\, C_2)\,, \qquad
\Gc = (C_3\,, \,C_4)\,,\label{222}
\end{equation}
where
\begin{equation}
(C_i\,,\,C_j) \equiv C_i\cos{Z}+C_j\sin{Z}\,,\label{DefCC}
\end{equation}
$C_1,\, C_2, \, C_3$ and $C_4$ are complex integration constants and
\begin{equation}
Z = 2\phib \sqrt{2c^2}\,\bpm
\end{equation}
(note that $\kpp= \ppp Z$). Substituting \eqref{222} into \eqref{221} we
find two relations between the integration constants
\begin{equation}
C_3 = -\frac12\,\sqrt{2c^2}\,C_2\,, \quad C_4 = \frac12\,\sqrt{2c^2}\,C_1\,.
\end{equation}
Taking into account the definition \eqref{def1} and \eqref{def2}, as
well as the property that harmonic integrals of the full harmonic
derivatives are vanishing, we also find
\begin{align}\label{I1I2}
\int\,du\, (C_1\,,\,C_2)=&0, &
\int\,du\, (-C_2\,,\,C_1)=&\frac4{\sqrt{2c^2}}\,.
\end{align}
To compute these integrals, we start by noting that
\begin{equation}
\bpp\bmm - (\bpm)^2 = \lambda \equiv \frac12\,b^2\,,
\qquad b^2 = b^{ik}b_{ik}\,. \label{229}
\end{equation}
It is then easy to show that
\begin{equation}
(\bpm)^{2k+1} = \partial^{++}\xi^{--}, \quad (\bpm)^{2k}
= \partial^{++}\chi^{--}
+(-1)^k\frac{1}{2k+1}\lambda^k\,,  \label{evenoddb}
\end{equation}
where $\xi^{--}$ and $\chi^{--}$ are some harmonic functions whose
precise form is of no relevance for computing integrals \eqref{I1I2}.
From eqs.  \eqref{evenoddb} it follows that the harmonic integral of any
odd power of $\bpm$ is vanishing, whence, e.g., $\int du \sin Z = 0\,$.
In Appendix \ref{harmints} we give a list of relevant non-vanishing
harmonic integrals.  Consulting it, we solve for the constants in
\eqref{I1I2}
\begin{equation}
C_1=0, \qquad\qquad C_2 = - \frac{4}{\sqrt{2 c^2\,}}\;\frac{X}{\sinh{X}}\,,
\end{equation}
where
\begin{equation}
X = 2\bar\phi\,\sqrt{c^2b^2}\,.   \label{DefX}
\end{equation}
The final form of the solution for $\Fc$ and $\Gc$ is
\begin{equation}
\Fc= \partial^{++}F^{--}=
-\frac4{\sqrt{2 c^2\,}}\left( 0\,,\,\frac{X}{\sinh{X}}\,\right) , \quad
\Gc= \partial^{++}B^{--} + 2 =
-2\;\left(\frac{ X}{\sinh{X}}\,,\,0\right) ,
\end{equation}
For our purpose of finding the closed form of $\delta A_{\alpha\alphad}$
there is no need to compute $B^{--}$ and $F^{--}$.  Indeed, the
coefficient of $\partial_{\alpha\alphad}a$ in \eqref{211},
\begin{equation}
L^\alpha_\beta  =
\delta^\alpha_\beta \left(1 - \phib \,c^2\,F^{--}\bpp\right)
+ \bpm c^\alpha_\beta + 2\,c^\alpha_\beta\,
\phib\,B^{--}\bpp +\frac12\ppp B^{--\;\alpha}_{1\beta}\,, \label{Lmat}
\end{equation}
is required not to depend on harmonics. Multiplying  \eqref{Lmat} by
$\int du\,1 = 1$ and integrating by parts with respect to the harmonic
derivative $\ppp$, we can represent this coefficient  as the
following harmonic integral
\begin{equation}
L^\alpha_\beta =
\int du \left[\delta^\alpha_\beta
\left(1 + 2\phib^2\,c^2\, \Gc\,\bpp\bmm\right)
+2\,c^\alpha_\beta\,\phib^2\,c^2\,\Fc\,\bpp\bmm\right].
\label{228}
\end{equation}
Keeping in mind \eqref{229} and the first of eqs. \eqref{evenoddb}, the
traceless part in \eqref{228} can be shown to vanish and the computation
of $L^\alpha_\beta$ is reduced to computing the harmonic integral
\begin{equation}
J = \int du (\bpp\bmm)\Gc\,.
\end{equation}
This is easy to perform using \eqref{229} and the formulas of Appendix
\ref{harmints}:
\begin{equation}
J = \frac{1}{2\phib^2c^2}\left(X\coth X - 1 \right)
= 2b^2 \left(\frac{X\coth X - 1}{X}\right).
\end{equation}
Finally, we obtain
\begin{equation}
L^\alpha_\beta = \delta^\alpha_\beta\,X\coth X\,.
\end{equation}
Thus the variation $\delta\;A_{\alpha\alphad}$ proves to have the very
simple form
\begin{equation}
\delta A_{\alpha\alphad} = \partial_{\alpha\alphad}a\,X\coth X \,.
\end{equation}
Note that $X\coth X$ is a well behaved function having the proper
undeformed limit $X\longrightarrow 0$,
\begin{equation}
\delta A_{\alpha\alphad} = \partial_{\alpha\alphad}a\left( 1 + \frac{X^2}{3}
-\frac{X^4}{45} + \ldots \right).
\end{equation}

\subsection{Gauge transformation of $\phi$}
The full gauge variation of $\phi$ is a sum of $\delta_0\phi$, eq.
\eqref{delta0gauge}, and the part computed from \eqref{deltaV} with
$\Lambda$ being substituted by $\Delta\Lambda$ \eqref{DeltaLambda}. We
have
\begin{eqnarray}
\delta \phi = \partial_{\alpha\alphad}a\,A^{\alphad}_\beta
\left[\partial^{++} G^{--\beta\alpha}
+ 4 \hat{C}^{+-\beta\alpha}
+2\,B^{--\;\alpha}_{1\rho} \hat{C}^{++\rho\beta}\right] \equiv
\partial_{\alpha\alphad}a\,A^{\alphad}_\beta\, G^{\beta\alpha}\,.
\end{eqnarray}
The r.h.s. of this transformation must be independent of harmonics as
stated in \eqref{gaugeWZcondb}.  This amounts to the condition
\begin{equation}
(\partial^{++})^2G^{--\;\alpha}_\beta
+2\hat{C}^{++\rho}_\beta\,\partial^{++}B^{--\;\alpha}_{1\rho} = 0\,.
\end{equation}
It can be easily solved for the unknown $G^{--\beta\alpha}$. However, it
is not necessary to know this solution for finding $\delta\phi$, since
\begin{equation}
\delta \phi =\partial_{\alpha\alphad}a\,A^{\alphad}_\beta\,
\int du \,G^{\beta\alpha} \label{235}
\end{equation}
and the problem is reduced to the computation of the harmonic integral
in \eqref{235}. For the product ansatz \eqref{Ccb} this integral, modulo
a total harmonic derivative in the integrand, is given by
\begin{equation}
\int du\, G^{\beta\alpha} = -\int du
\left( \epsilon^{\alpha\beta}\,c^2\, \bpm\,\Fc
+2\,c^{(\alpha\beta)}\,\bpm\,\Gc\right). \label{236}
\end{equation}
The second term under the integral in \eqref{236} is a total harmonic
derivative in virtue of the first relation in \eqref{evenoddb} and is
therefore vanishing, whence
\begin{equation}
\delta \phi = 2\,\sqrt{c^2b^2}\,
(\partial_{\alpha\alphad}a\,A^{\alpha\alphad})\,
\frac{1-X\,\coth X}{X}\,.
\label{239}
\end{equation}
Note that the r.h.s. of \eqref{239} contains only integer powers of
$c^2b^2$, as can be seen from the definition \eqref{DefX} and the fact
that the function of $X$ in \eqref{239} contains only odd powers of
$X\,$.

\subsection{The full set of  QNS-deformed gauge variations}
\label{minimalmap}
Proceeding in the same way as above, one can find the full set of
QNS-deformed gauge transformations laws for the $\Nc=(1,1)$ vector
multiplet in WZ gauge. Here is the list of them:
\begin{align}\label{gtoriginal}
\delta\,\phib= &0\,, \qquad \delta\Psib^k_{\alphad}=0\,, \\
\delta\,A_{\alpha\alphad}=& X\coth X \partial_{\alpha\alphad}a\, ,\\
\delta\,\phi= &2\sqrt{c^2\,b^2}\left( \frac{1-X\coth X}{X}\right)
A^{\alpha\alphad}\partial_{\alpha\alphad}a\, ,\\\nonumber
\delta\Psi^i_\alpha=&\Biggl\lbrace
\left[\frac{4X^2(X\coth X-1)}{X^2+\sinh^2X-X\sinh2X}
\right] b^{ij}c_{\alpha\beta}\\
&-\sqrt{c^2b^2}\,
\left[\frac{4X\cosh^2X-2X^2(\coth X+X)-\sinh2X }
{X^2+\sinh^2X-X\sinh2X}\right]\ve^{ij}\ve_{\alpha\beta}
\Biggr\rbrace \Psib_{j\alphad}\,\partial^{\beta\alphad}a\,.\\
\delta D_{ij}=& 2\iim
b_{ij}c^{\alpha\beta}\partial_{\alpha\alphad}\phib\,
\partial^{\alphad}_\beta a\,.
\end{align}

Having the explicit form of the deformed gauge transformations, one can
deduce a minimal Seiberg-Witten-like map (SW map) which puts these
transformations into the standard undeformed form
\begin{align}
A_{\alpha\alphad}=&\,\widetilde{A}_{\alpha\alphad}\, X\coth X\, ,\\
\phi=&\,\widetilde{\phi}
+\widetilde{A}^2\;\sqrt{c^2\,b^2}\;X\coth X
\left( \frac{1-X\coth X}{X}\right) ,\\\nonumber
\Psi^{i\alpha}=&\widetilde{\Psi}^{i\alpha}
+\Biggl\lbrace
\left[\frac{4X^2(X\coth X-1)}{X^2+\sinh^2X-X\sinh2X}
\right] b^{ij}c_{\alpha\beta}\\
&-\sqrt{c^2b^2}\,
\left[\frac{4X\cosh^2X-2X^2(\coth X+X)-\sinh2X }
{X^2+\sinh^2X-X\sinh2X}\right]\ve^{ij}\ve_{\alpha\beta}
\Biggr\rbrace \Psib_{j\alphad}\,\widetilde{A}^{\beta\alphad}\,.\\
D_{ij}=& \widetilde{D}_{ij}+2\iim
b_{ij}c^{\alpha\beta}\partial_{\alpha\alphad}\phib\,
\widetilde{A}^{\alphad}_\beta\,.
\end{align}
For the fields with ``tilde'' we obtain the standard transformations
\begin{align}
&\delta \widetilde{A}_{\alpha\alphad}=
\partial_{{\alpha\alphad}}a\,,
\qquad \delta \widetilde\phi=0\,,\qquad
\delta\widetilde{D}^{ij}=0\,,\qquad
\delta\widetilde{\Psi}^{k}_{\alpha}=0\,.
\end{align}
The gauge field strength
$F_{\alpha\beta}=2\iim\partial_{(\alpha\alphad}A^{\alphad}_{\beta)}$
which is non-covariant with respect to the deformed transformations is
redefined under the transformation $A_{\alpha\alphad} \rightarrow
\widetilde{A}_{\alpha\alphad}$ as
\begin{equation}
F_{\alpha\beta}=\widetilde{F}_{\alpha\beta}
X\coth X+4\iim\sqrt{b^2\,c^2}\widetilde{A}_{(\beta\alphad}
\partial^{\alphad}_{\alpha)}\phib
\left(\coth X - \frac{X}{\sinh^2{X}}\right), \label{FF}
\end{equation}
where
$\widetilde{F}_{\alpha\beta}=2\iim\partial_{(\alpha\alphad}
\widetilde{A}^{\alphad}_{\beta)}\,$.
Since $\widetilde{F}_{\alpha\beta}$ is manifestly gauge invariant, it is
easy to derive from \eqref{FF} what is the deformed analog of this field
strength.

\section{The QNS-deformed action}
In this section we calculate the $\Nc=(1,0)$ gauge invariant
action in components. We concentrate on the bosonic limit of the action.
The full supersymmetric action with all the fermionic fields included,
as well as the full set of unbroken $\Nc=(1,0)$ supersymmetry
transformation laws,\footnote{In fact, it is of no actual necessity to
explicitly know these transformations, since our procedure of deriving
the action is manifestly $\Nc=(1,0)$ supersymmetric by construction.}
will be presented in a forthcoming paper \cite{In prep}.

The QNS-deformed action  for the $\Nc=(1,1)$ U(1) gauge theory in
harmonic superspace \cite{Galperin:2001uw}, in the form most appropriate
for our purposes, is written in the same way as in the QS-deformed case
\cite{Ferrara:2004zv}
\begin{equation}
    S=\frac14\int d^4x_L\,d^4\theta\,du\,\Wc\star\Wc=
    \frac14\int d^4x\,d^4\theta\,du\,\Wc^2\,.
\end{equation}
Here $\Wc$ is the covariant superfield strength
\begin{equation}\label{Wc}
    \Wc=-\frac14(\bar{D}^+)^2 \Vmm\equiv
    \Ac+\thetab^+_\alphad\tau^{-\alphad}+(\thetabp)^2\tau^{--}\,,
\end{equation}
and $\Vmm$ is the non-analytic harmonic connection related to $\V$ by the
harmonic flatness condition
\begin{equation}\label{flateqVmm}
\Dpp V^{--}-\Dmm\V+\left[ \V\,,\, V^{--}\right]_\star = 0\,.
\end{equation}
In \eqref{Wc} we have used the general expansion of $\Vmm$ in terms of
chiral superfield components (depending only on $x_L^{\alpha\dot\alpha},\,
\theta^{\pm}_\alpha,\, u^{\pm i}$)
\begin{align}
\label{Vmm}
\nonumber\Vmm=\vmm+\thetab^+_\alphad v^{(-3)\;\alphad}
&+\thetab^-_\alphad v^{-\;\alphad}
+(\thetabm)^2 \Ac+(\thetabp\thetabm)\varphi^{--}
+\thetabm^\alphad\thetabp^\betad\varphi^{--}_{\alphad\betad}
+(\thetabp)^2 v^{(-4)}\\
&+(\thetabm)^2\thetab^+_\alphad \tau^{-\;\alphad}
+(\thetabp)^2\thetab^-_\alphad \tau^{(-3)\;\alphad}
+(\thetabp)^2(\thetabm)^2\tau^{--}\,.
\end{align}

The whole effect of the considered deformation in the above action comes
from the structure of $\Wc$ due to the presence of the star commutator
in the equation \eqref{flateqVmm} defining $\Vmm$. As a consequence of
the latter, \eqref{Wc} satisfies the condition
\begin{equation}\label{flateqWc}
\Dpp\Wc+\left[ \V\,,\,\Wc\right]_\star = 0\,,
\end{equation}
which amounts to the following equations for the chiral coefficients in
\eqref{Wc}
\begin{subequations}
\begin{align}
&\label{nablappAc}\nablapp\Ac=0\,,\\[2mm]
&\label{nablapptaum}\nablapp\tau^{-\alphad}
+\left[ v^{+\alphad},\Ac \right]_\star=0\,,\\[2mm]
&\label{nablapptaumm}\nablapp\tau^{--}
-\frac12\left\{ v^+_\alphad, \tau^{-\alphad} \right\}_\star
+\left[ v,\Ac \right]_\star=0\,,
\end{align}
\end{subequations}
where
\begin{equation}
\label{nablapp}
\nablapp={\Dpp} +\left[\vpp,\right]_\star
\end{equation}
and $v^{++} = (\theta^+)^2 \phib$, $v^+_\alphad$ and $v$ are defined in
\eqref{vp}, \eqref{v}. The Q-deformed commutator in \eqref{nablapp}, for
a  general chiral superfield $\Phi(x_L, \theta^\pm_\alpha)\,$
(irrespective of the Grassmann parity of the latter), reads
\begin{equation}
\left[\vpp,\Phi\right]_\star= -2\partial_{+\alpha}\vpp\partial_{+\beta}\Phi\,
\hat{C}^{++ \alpha\beta}-2\partial_{+\alpha}\vpp\partial_{-\beta}\Phi\,
\hat{C}^{+- \alpha\beta}
-2\,I\,\partial_{+\alpha}\vpp\partial_{-\beta}\Phi\,\ve^{\alpha\beta}\,.
\end{equation}
Then for the product ansatz \eqref{Ccb}, $\nablapp\Phi$ becomes
\begin{equation}
\nablapp\Phi=\left[\ppp -\left(\ve^{\alpha\beta}
+4\phib\bpm c^{\alpha\beta}\right)
\theta^+_\alpha\partial_{-\beta}
-4\phib\bpp\,c^{\alpha\beta}
\theta^+_\alpha\partial_{+\beta}\right]\Phi\,.
\end{equation}

Like in the QS-case \cite{Ferrara:2004zv}, using  eqs.
\eqref{nablapptaum}, \eqref{nablapptaumm} and the following additional
equations implied by \eqref{flateqVmm} for the other chiral coefficients
in the expansion \eqref{Vmm}
\begin{subequations}
\begin{align}
&\label{nablappvm}\nablapp v^{-\;\alphad}-v^{+\;\alphad}=0\,,\\[3mm]
&\label{nablappvarphimm}\nablapp\varphi^{--}+2(\Ac-v)
+\frac12\left\lbrace v^{+\;\alphad},v^-_\alphad \right\rbrace_\star =0\,,
\end{align}
\end{subequations}
it is straightforward to find the explicit form of $\tau^{-\dot\alpha},
\tau^{--}$ and to show that the only superfield that contributes to the
action is $\Ac\,$. Thus the invariant action is reduced to
\begin{equation}\label{actionAc}
    S=\frac14\int d^4x\,d^4\theta\,du\,\Ac^2\,,
\end{equation}
and it remains to calculate the superfield $\Ac\,$.  This can be
accomplished using eqs. \eqref{nablappAc}, \eqref{nablappvm} and
\eqref{nablappvarphimm}.  As a first step, we substitute in
\eqref{actionAc} the component expansion of $\Ac$
\begin{align}\label{Aexpansion}
\nonumber \Ac=\Ac_1+\thetam^{\alpha}\Ac^+_{2\;\alpha}
&+\thetap^{\alpha}\Ac^-_{3\;\alpha}
+(\thetam)^2\Ac^{++}_4 +(\thetam\thetap)\Ac_5
+\thetam^\alpha\thetap^\beta \Ac_{6\;\alpha\beta}
+(\thetap)^2\Ac^{--}_7\\[2mm]
&+(\thetam)^2\thetap^{\alpha}\Ac^+_{8\;\alpha}
+(\thetap)^2\thetam^{\alpha}\Ac^-_{9\;\alpha}
+(\thetam)^2(\thetap)^2\Ac_{10}\,,
\end{align}
and integrate over Grassmann coordinates, which leads to
\begin{equation}
    S=\frac14\int d^4x_L\,du\,
    \left[
    2\Ac_1\Ac_{10}-(\Ac^+_2\Ac^-_9)-(\Ac^-_3\Ac^+_8)
    +2\Ac^{++}_4\Ac^{--}_7-\frac12\Ac_5^2-\frac14\Ac_6^2
    \right].
\end{equation}
As soon as we limit ourselves to the bosonic part of the action, the terms
$\Ac^+_{2\,\alpha}, \Ac^-_{3\,\alpha},\Ac^+_{8\,\alpha}$ and
$\Ac^-_{9\,\alpha}$ can be discarded. Thus the bosonic action
we are searching for takes the form
\begin{equation}\label{bosonicact:bacomp}
    S_{bos}=\frac14\int d^4x_L\,du\,
    \left[
    2\Ac_1\Ac_{10}
    +2\Ac^{++}_4\Ac^{--}_7-\frac12\Ac_5^2-\frac14\Ac_6^2
    \right].
\end{equation}

\subsection{Determining $\Ac$}
By substituting the $\theta$ expansion \eqref{Aexpansion}, with all
fermionic components omitted, into the superfield equation
\eqref{nablappAc}, we obtain the following equations for the quantities
entering the bosonic action \eqref{bosonicact:bacomp}:
\begin{subequations}
\begin{align}
&\label{a1}\ppp\Ac_1=0\,,\\[3mm]
&\label{a4}\ppp\Ac^{++}_4=0\,,\\[3mm]
&\label{a5}\ppp\Ac_5+2\Ac^{++}_4
-2\phib\bpp\;c^{\alpha\beta}\;\Ac_{6\alpha\beta}=0\,,\\[3mm]
&\label{a6}\ppp\Ac_{6\alpha\beta}+4\phib(2\bpm \Ac^{++}_4
+\bpp\Ac_5)\;c_{\alpha\beta}
+4\bpp\phib\Ac_{6(\alpha\gamma}c^\gamma_{\beta)}=0\,,\\[3mm]
&\label{a7}\ppp\Ac^{--}_7+\Ac_5
+2\phib\bpm c^{\alpha\beta}\Ac_{6\alpha\beta}=0\,,\\[3mm]
&\label{a10}\ppp\Ac_{10}=0\,.
\end{align}
\label{nablappAcComp}
\end{subequations}

It immediately follows from \eqref{a1}, \eqref{a4} and \eqref{a10} that
$\Ac_1$ and $\Ac_{10}$ are independent of harmonics and that
$\Ac^{++}_4$ is of the form $\Ac^{++}_4=\Ac^{ij}_4u^+_iu^+_j$, with
$\Ac^{ij}_4$ being independent of harmonics. It is also obvious  that
the equations \eqref{nablappAcComp}, being homogeneous, can determine
the components of $\Ac$ only up to some integration constants.  These
constants should be fixed from eqs. \eqref{nablappvarphimm} and
\eqref{nablappvm}. To accomplish this, we first  have to pass to
components in $v^{-\;\alphad}$ and $\vpmm$ as in \eqref{Aexpansion} and
substitute the relevant $\theta$ expansions (with all fermions omitted)
into eqs. \eqref{nablappvarphimm} and \eqref{nablappvm}. After that we
should solve the corresponding harmonic equations for the components.
Omitting details, we obtain the following solutions for $\Ac_1$ and
$\Ac_{10}$:
\begin{eqnarray}
\Ac_1 &=& \phi + \frac12\phib^{-1}\left( 1-\frac{\tanh X}{X}\right)A^2
+(b^2\, c^2)^{3/2}
\tanh X \partial_{\alpha\alphad}\phib\partial^{\alpha\alphad}\phib\,, \nn
\Ac_{10}&=&-\square\phib\,, \qquad
A^2 \equiv A^{\alpha\dot\alpha}A_{\alpha\dot\alpha}\,.
\label{A1A10}
\end{eqnarray}

Solving the rest of eqs. \eqref{nablappAcComp} is not so easy, though
things are simplified by observing that we do not need to calculate
$\Ac_7^{--}$.  Indeed, making use of the relation $\Ac^{++}_4 =
\partial^{++}\Ac^{+-}_4, \Ac^{+-}_4 = \Ac_4^{(ij)}u^+_iu^-_j$ and
integrating by parts in the term
\begin{equation}
\int\,d^4x\,du\, \frac12\Ac^{++}_4\,\Ac_7^{--}=
-\int\,d^4x\,du\,\frac12 \Ac^{+-}_4\,\ppp\Ac_7^{--}\,,\label{SimplF}
\end{equation}
in \eqref{bosonicact:bacomp}, we can eliminate $\ppp\Ac_7^{--}$ with the
help of eq. \eqref{a7}. As a result, the remaining part of the action
will involve only $\Ac^{ij}_4$, $\Ac_5$ and $\Ac_{6\alpha\beta}\,$.  Our
strategy for finding explicit expressions for these quantities as
solutions of the appropriate component harmonic equations is as follows.
Firstly we find series solutions to few first orders in the deformation
parameters, with taking account of the component equations comprised by
\eqref{nablappvarphimm}.  Such a perturbative solution suggests a
particular ansatz for the sought component fields, which finally
provides an exact solution reproducing the known terms in the series
solution.  In this way we arrive at the following ansatz
\begin{equation}
\begin{aligned}
    \Ac_{6\alpha\beta}&=g_1\, F_{\alpha\beta}+g_2\,c_{\alpha\beta}
    +g_3\,c_{(\alpha}^\gamma F_{\gamma\beta)}
    +g_4\,A_{(\alpha\alphad}\partial^\alphad_{\beta)}\phib
    +g_5\,c_{(\alpha}^\gamma
    A_{(\gamma\alphad}\partial^\alphad_{\beta))}\phib\,,\\
    \varphi_{6\alpha\beta}^{--}&=h^{--}_1\,F_{\alpha\beta}
    +h^{--}_2\,c_{\alpha\beta}
    +h^{--}_3\,c_{(\alpha}^\gamma F_{\gamma\beta)}
    +h^{--}_4\,A_{(\alpha\alphad}
    \partial^\alphad_{\beta)}\phib +h^{--}_5\,
    c_{(\alpha}^\gamma A_{(\gamma\alphad}
    \partial^\alphad_{\beta))}\phib\,,\\
    \Ac^{ij}_4&=\alpha_1\,D^{ij}
    +\alpha_2\,b^{ij}+\alpha_2\,b^{(ik}D^{j)}_k\,.
\end{aligned}
\label{bosonicact:SymAnsatz}
\end{equation}
The functions $g_i$, $h^{--}_i$ can depend on $\phib, b^{ij}$,
$c_{\alpha\beta}$ and harmonics. Similarly, $\alpha_i$ can depend only
on $\phib, b^{ij}$ and $c_{\alpha\beta}$; these functions are harmonic-
independent. To find all these functions one actually needs eq.
\eqref{a6} (which amounts to several coupled equations after
substituting the above ansatz for $\Ac_{6\alpha\beta}$) and those
component equations which appear as the coefficients of the monomials
$(\theta^-)^2, (\theta^{-\alpha}\theta^{+\beta})$ and $(\theta^+)^2\,$
in \eqref{nablappvarphimm}.  Once the unknowns in
\eqref{bosonicact:SymAnsatz} are found, the function $\Ac_5$ can be
computed from \eqref{a5}, without assuming beforehand any ansatz for it.
Skipping details of calculations and introducing the short-hand
notation
\begin{eqnarray}
&& c\cdot F=c^{\alpha\beta}F_{\alpha\beta}\,, \quad
b\cdot D=b_{ij}D^{ij}\,, \quad
c\cdot A\partial\phib =c^{\alpha\beta}\,
A_{(\alpha\alphad}\partial^\alphad_{\beta)}\phib\,, \nonumber \\
&& A\cdot \partial\phib =
A_{\alpha\alphad}\partial^{\alpha\alphad}\phib\,, \quad
F\cdot A\partial\phib =F^{\alpha\beta}\,
A_{(\alpha\alphad}\partial^\alphad_{\beta)}\phib\,, \nonumber
\end{eqnarray}
the final expressions for the remaining building blocks of the bosonic
action are
\begin{eqnarray}
\Ac^{ij}_4 &=& \frac{\sigma\,b^{ij}}{\cosh^3X}
+\frac{D^{ij}}{\cosh^2X}\,, \label{resultA4}\\
\Ac_5 &=& -\frac{2\sigma\,\bpm}{\cosh^3X}+\frac{\sqrt{2b^2}}{X^2\cosh^3X}
\Bigl(0\,,\,-2\iim\cosh^2X(\cosh X\sinh X-1)(c\cdot A\partial\phib)\nn
&&-\,X^2\sinh X\sigma+
\phib\cosh^3X\sinh X (c\cdot F)\Bigl)+\frac{D^{+-}}{\cosh X}(-2\,,\,0)\nn
&&+\,\frac{\phib\sqrt{2c^2}\bmm\, D^{++}}{(X^2+Z^2)}
\left[(0\,,\,-X\sinh X)-Z+(\cosh X\,,\, 0)Z \right],
\label{resultA5}
\end{eqnarray}
\begin{subequations}\label{resultA6}
\begin{align}
g_1=&\left( \frac{\sinh X}{X}\,,\,0\right),\\[3mm]
g_2=&-\frac{2\phib\,b^2\left[(X^2+Z^2)\sigma
+8\phib^2c^2\cosh X\bmm\,D^{++}\right]}{X^2\cosh^3X(X^2+Z^2)}
\left[ (0\,,\,2\cosh X)Z+(2X\sinh X\,,\,0)\right] \cr&+
\sqrt{\frac{2}{c^2}}\left(0\,,\, \frac{2D^{+-}}{\cosh X}
+\frac{2\bpm\sigma}{\cosh^2X}\right),\\[3mm]
g_3=&-\sqrt{\frac2{c^2}}\left( 0\,,\, \frac{\sinh X}{X}\right),\\[3mm]
g_4=&\frac{2\iim\sqrt{b^2c^2}}{X^2\cosh X}
\left(X-\cosh X\sinh X\,,\, 0\right),\\[3mm]
g_5=&\frac{2\iim\sqrt{b^2c^2}}{X^2\cosh X}
\left(0\,,\,\cosh X\sinh X-X\right).
\end{align}
\end{subequations}
Here
\begin{equation}
\sigma= \frac{\sinh X}{X}\left[-2\iim (c\cdot A\partial\phib)\,
\frac{\cosh X\sinh X -2X}{X}
+(c\cdot F) \phib \,\frac{\cosh X\sinh X}{X}\right]
\end{equation}
and we used the definition \eqref{DefCC}.

All these expressions meet the criterion of regularity in $b^2$, $c^2$
and $\bar\phi$, as can be checked on their closer inspection. In
particular, functions multiplied by $1/(X^2+Z^2)$ in the expressions for
$\Ac_5$ and $g_2$ fulfill the consistency and regularity conditions which
mean that they are in fact of the form $(X^2+Z^2) F(X,Z)\,$, where
$F(X,Z)$ is regular.  This regularity can be seen most clearly by
representing
\begin{eqnarray}
&&\frac{1}{X^2+Z^2}
\left[(0\,,\,-X\sinh X)-Z+(\cosh X\,,\, 0)Z \right]\nn
&&= \frac{\iim}{2}\left( \frac{\cosh (X+\iim Z) -1}{X+\iim Z}
-\frac{\cosh (X-\iim Z) -1}{X-\iim Z}\right), \nn
&&\frac{1}{X^2+Z^2}
\left[ (2X\sinh X\,,\,0) + (0\,,\,2\cosh X)Z\right] =
\frac{\sinh (X+\iim Z)}{X+\iim Z}+ \frac{\sinh (X-\iim Z)}{X-\iim Z}\,.
\end{eqnarray}
As an example, we quote the first terms in the series expansion of
$\Ac^{ij}_4$ and $\Ac_5$
\begin{equation}
\Ac^{ij}_4=D^{ij}+b^{ij}[2\iim(c\cdot A\partial\phib)+\phib (c\cdot F)]
+\cdots\,, \qquad
\Ac_5=-2D^{+-}-4\iim\bpm(c\cdot A\partial\phib)+\cdots\, .
\end{equation}

Substituting the expressions \eqref{A1A10}, \eqref{resultA4},
\eqref{resultA5} and the expression for $\Ac_{6\alpha\beta}$ with the
functions $g_i$ given by \eqref{resultA6} into \eqref{bosonicact:bacomp}
(with taking into account \eqref{SimplF}) and doing the harmonic
integral we obtain
\begin{align}
    S_{bos}=&\int\,d^4x \Biggl\{-\frac12\left[\phi
    +\frac12 \phib^{-1}\left( 1-\frac{\tanh X}{X}\right)A^2
    +(b^2\, c^2)^{3/2} \tanh X
    \partial_{\alpha\alphad}\phib\partial^{\alpha\alphad}\phib\right]
    \square\phib \nonumber\\
    &+\frac14\frac{D^2}{\cosh^2 X}-\frac1{16}F^2\frac{\sinh^2 X}{X^2}
    +\frac12\phib(b\cdot D)(c\cdot F)\frac{\tanh^2 X}{X^2}
    +\!\frac14b^2(c\cdot F)^2\phib^2\frac{\sinh^4 X}{X^4\cosh^2 X}\nonumber\\
    &-\iim\left[(b\cdot D)(c\cdot A\partial\phib)\frac{\tanh X}X
    +b^2(c\cdot A\partial\phib)(c\cdot F)\phib
    \frac{\sinh^3 X}{X^3\cosh X}\right]
    \left(\frac{\tanh X}X-\frac2{\cosh^2 X}\right)\nonumber\\
    &-(c\cdot A\partial\phib)^2b^2\frac{\sinh^2 X}{X^2}
    \left(\frac{\tanh X}X-\frac2{\cosh^2 X}\right)^2
    \!\!+\frac{(A\partial\phib)^2b^2c^2}{\cosh^2 X}
    \left( \frac{\cosh X\sinh X-X}{X^2} \right)^2\nonumber\\
    &+\iim\sqrt{b^2c^2}\frac{(F\cdot A\partial\phib)}2\frac{\tanh X}X
    \left( \frac{\cosh X\sinh X-X}{X^2} \right)\Biggl\}.\label{Act1}
\end{align}
Through the minimal SW map (see subsection \ref{minimalmap}) this action
is simplified to
\begin{align}
    S_{bos}=\int\,d^4x\,\Bigl[&-\frac{1}{2}\tilde{\phi}\square\phib
    -\frac12 (b^2\, c^2)^{3/2} \tanh X \partial_{\alpha\alphad}\phib
    \partial^{\alpha\alphad}\phib\square\phib
    +\frac14 \frac{\widetilde{D}^2}{\cosh^2{X}}\nonumber\\
    &-\frac1{16}\widetilde{F}^2\cosh^2X
    +\frac14b^2(c\cdot\widetilde{F})^2\phib^2\frac{\sinh^2 X}{X^2}
    +\frac12\phib(b\cdot\widetilde{D})
    (c\cdot\widetilde{F})\frac{\tanh{X}}X\Bigr].
\end{align}
This action is invariant under the standard abelian gauge
transformations. Turning off the deformation parameters we are left with
the usual bosonic sector of the undeformed action. Performing the
further field redefinition
\begin{subequations}
\begin{align}
    d^{ij}&=\frac1{\cosh^2 X}\widetilde{D}^{ij}
    +\phib(c\cdot\widetilde{F})b^{ij}\frac{\tanh X}X\,,\\
    \varphi&=\frac1{\cosh^2 X}\left[ \tilde\phi
    +(b^2c^2)^{3/2}(\partial\phib)^2\tanh X \right],
\end{align}\label{redef}
\end{subequations}
the bosonic action can be transformed to the most simple form
\begin{equation}\label{finalaction}
S_{bos}=\int\,d^4x\; \cosh^2{X} \left[- \frac12\varphi\square\phib
+ \frac14 d^{ij}d_{ij}
-\frac1{16}\tilde{F}^{\alpha\beta}\tilde{F}_{\alpha\beta} \right].
\end{equation}
From this expression it is obvious that we cannot disentangle the
interaction between the gauge field and $\phib$ by any field
redefinition. This is similar to the singlet case
\cite{Ferrara:2004zv,Araki:2004mq}, where the scalar factor $(1 +
4I\bar\phi)^2$ appears instead of $\cosh^2 X\,$. Note that the bosonic
action \eqref{finalaction} involves only squares $c^2$ and $b^2$, so it
preserves ``Lorentz'' $Spin(4)$=
SU(2)$_{\text{L}}\times$SU(2)$_{\text{R}}$ symmetry and SU(2) R-symmetry
as in the singlet case, despite the fact that the deformation matrix
\eqref{Ccb} breaks both these symmetries down to
U(1)$_{\text{L}}\times$SU(2)$_{\text{R}}$ and U(1). This property is
similar to what happens in the deformed Euclidean $\Nc = (1/2,1/2)$
Wess-Zumino model where the deformation parameter $C^{\alpha\beta}$ also
appears through its Lorentz-invariant square \cite{Seiberg:2003yz}.
Note, however, that the fermionic completion of \eqref{finalaction}
explicitly includes both $c^{\alpha\beta}$ and $b^{ik}$ \cite{In prep},
so these two symmetries are broken in the total action.  This feature
matches with the fact that Lorentz symmetry is broken in the action of
deformed $\Nc = (1/2,1/2)$ gauge theory, and also due to some fermionic
terms \cite{Seiberg:2003yz}.

\subsection{Limiting cases}
Let us expand the action \eqref{Act1} up to the first order in the
deformation parameters $b_{ij}$ and $c^{\alpha\beta}\,$ to compare it
with the results of \cite{Araki:2004mq, Araki:2004cv, Araki:2005pc}.
In this approximation the action reads
\begin{equation}\label{action-1er-orden}
S_{bos}=\int d^4x_L\; \left[-\frac12\phi\square\phib
+\frac14 D^2+\iim c^{\alpha\beta}A_{\alpha\alphad}
\partial_{\beta}^{\alphad}\phib\;
D^{ij}b_{ij}-\frac1{16} F^2
+\frac12 \phib D^{ij}b_{ij}
c^{\alpha\beta}F_{\alpha\beta}\right].
\end{equation}
It can be checked that \eqref{action-1er-orden} coincides with the
first-order action of references \cite{Araki:2004mq, Araki:2005pc}
upon substituting there the product ansatz \eqref{Ccb}. The gauge
transformations laws \eqref{gtoriginal} in this case reduce to
\begin{subequations}
\begin{align}
&\delta A_{\alpha\alphad}=\partial_{\alpha\alphad} a\,,\\
&\delta \phi=0\,,\\
&\delta \Psi^i_\alpha
=-\frac43b^{ij}c_{\alpha\beta}\Psib_{j\alphad}
\partial^{\beta\alphad} a\,,\\
&\delta D_{ij}=2\iim b_{ij} c^{\alpha\beta}
\partial_{\alpha\alphad}\phib
\partial_{\beta}^{\alphad} a\,.
\end{align}
\end{subequations}
These laws also precisely match with those given in
\cite{Araki:2004mq, Araki:2005pc}.

It would be also interesting to study the particular case of our results
corresponding to the choice
\begin{equation}\label{3/2}
c^{\alpha\beta}\neq0\, \;\;b_{11}\neq 0\,,\;\;
b_{12}=b_{22}=0 \rightarrow b^2 = 0\,.
\end{equation}
This degenerate choice preserves $\Nc=(1,1/2)$ supersymmetry
\cite{Ivanov:2003te,Araki:2005pc}.  First we focus on the non-trivial
gauge transformations \eqref{gtoriginal}, which for this choice are
reduced to
\begin{align}\label{gt3/2}
&\delta A_{\alpha\alphad}=
\partial_{\alpha\alphad}a\,,
\qquad\delta\phi=0\,, \qquad \delta\Psi^1_{\alpha}=0\,,\crcr
&\delta\Psi^2_{\alpha}=\frac43 b_{11}c_{\alpha\beta}
\Psib^1_\alphad\, \partial^{\beta\alphad}a, 
\qquad \delta D_{11}=2\iim b_{11} c^{\alpha\beta}
\partial_{(\alpha\alphad}\phib
\partial_{\beta)}^{\alphad}a\,,
\qquad \delta D_{12}= \delta D_{22}=0\,.
\end{align}
Then we look at the action \eqref{action-1er-orden},
\begin{equation}
S_{bos}=\int d^4x_L \left[ -\frac12\phi\square\phib
-\frac1{16} F^{\alpha\beta}F_{\alpha\beta}
+\frac14 D^2+\iim D^{11}b_{11}\, c^{\alpha\beta}
\partial_{(\alpha\alphad}\phib A_{\beta)}^\alphad
+\frac12\phib D^{11} b_{11} c^{\alpha\beta}
F_{\alpha\beta}\right].
\end{equation}
Since in our case $b^2 = 0$, this action is actually the {\it exact\/}
form of \eqref{Act1} for the considered choice.  To compare our results
with those on \cite{Araki:2005pc} we have to use the minimal SW map for
the choice \eqref{3/2}, which is easy to calculate. In fact, the only
non-trivial transformation is
\begin{equation}
D_{11}=\widetilde{D}_{11}
-2\iim b_{11}c^{\alpha\beta}
\widetilde{A}_{(\alpha\alphad}
\partial_{\beta)}^\alphad\phib\,.
\end{equation}
Performing it  gives rise to the following action
\begin{equation} \label{action222}
S_{bos}=\int d^4x_L\; \left[ -\frac12\tilde{\phi}\square\phib
-\frac1{16} \widetilde{F}^{\alpha\beta}
\widetilde{F}_{\alpha\beta}+\frac{1}{4} \widetilde{D}^2
+\frac12\phib \widetilde{D}^{11}\,b_{11}
c^{\alpha\beta} \widetilde{F}_{\alpha\beta}\right].
\end{equation}
This action, as well as the gauge transformations \eqref{gt3/2},
coincide with the expressions for the gauge transformations and bosonic
sector of the action given in \cite{Araki:2005pc}, up to a constant
rescaling of the fields. Thus, proceeding from the ${\cal N}=(1,1)$ superfield
formalism, we have reproduced the results of \cite{Araki:2005pc} obtained
within the ${\cal N}=1$ superfield formalism.
It is interesting to note that the redefinitions \eqref{redef} are reduced to
\begin{align}
&\varphi= \tilde\phi\,,\crcr
&d^{11}={D}^{11}\,,\crcr
&d^{12}={D}^{12}\,,\crcr
&d^{22}={D}^{22}
+\phib\;c^{\alpha\beta}\widetilde{F}_{\alpha\beta}\;b^{22}\,,
\end{align}
and the action \eqref{action222} becomes undeformed in terms of the new
fields
\begin{equation}
S=\int\,d^4x\; \left[- \frac12\varphi\square\phib
+\frac14 d^{ij}d_{ij}
-\frac1{16}\widetilde{F}^{\alpha\beta}\widetilde{F}_{\alpha\beta} \right],
\end{equation}
which is obviously consistent with our result \eqref{finalaction}.

\section{Unbroken $\Nc=(1,0)$ supersymmetry}
As we pointed out in the beginning of Sect. 4, there is no actual need
to explicitly know the unbroken $\Nc=(1,0)$ supersymmetry
transformations, since the full action (with all fermionic terms taken
into account \cite{In prep}) is supersymmetric by construction.
Nevertheless, for completeness, here we explain how to derive the
$\Nc=(1,0)$ transformations in the WZ gauge in the present case and give
some examples of these transformations.

Unbroken supersymmetry is realized on $\V$ as
\begin{equation}\label{unbrokensusy}
\delta\V=\left( \epsilon^{+\alpha}\partial_{+\alpha}
+\epsilon^{-\alpha}\partial_{-\alpha}\right)\V
-\Dpp\Lambda_c-\left[ \V\, ,\,\Lambda_c \right]_\star\,,
\end{equation}
where the star bracket, like in the previous consideration, is defined
via the non-singlet Poisson structure with the deformation matrix
\eqref{Ccb} and $\Lambda_{c}$ is the compensating gauge parameter which
is necessary for preserving WZ gauge.

As in the case of deformed gauge transformations, for ensuring the
correct undeformed limit, the parameter $\Lambda_c$ should start with
the parameter $\L$ corresponding to the undeformed $\Nc=(1,0)$
supersymmetry.  Being expressed in chiral coordinates, $\L$ is
\cite{Ferrara:2004zv}
\begin{equation}
\L=\l+(\thetabp\l^-)+(\thetabp)^2\l^{--}\,,
\end{equation}
where
\begin{equation}
    \begin{aligned}
    \lambda_\epsilon&=2(\epsilonm\thetap)\phib\,,\\[3mm]
    \l^{-\alphad}&=
    4\iim(\epsilonm\thetap)\theta^-_\alpha\partial^{\alpha\alphad}\phib
    -2\epsilon^-_\alpha A^{\alpha\alphad}+
    4(\epsilonm\thetap)\Psibm^{\alphad}\,,\\[3mm]
    \l^{--}&=2(\epsilon^-\Psi^-)
    +2\iim\epsilonm^{\alpha}\thetam^{\beta}
    \partial^{\alphad}_{\beta}A_{\alpha\alphad}
    -2(\epsilonm\thetap)(\thetam)^2\square\phib\\[3mm]
    &\quad+4\iim(\epsilonm\thetap)\thetam^{\alpha}
    \partial_{\alpha\alphad}\Psibm^{\alphad}
    +2(\epsilonm\thetap)D^{--}\,.
    \end{aligned}
\end{equation}
Let us for a while drop the undeformed part of the $\Nc=(1,0)$ variation
of $\V$, which we denote by $\delta_0\V$ and which corresponds to
substituting $\L$ for $\Lambda_c$ in \eqref{unbrokensusy} and discarding
the last commutator term there.  We denote by $\check\delta\V$ the
lowest-order non-singlet part of the transformations coming from the
star commutator in \eqref{unbrokensusy} with $\L$ as the gauge
parameter. In terms of superfield components, this part has the form
\begin{equation}\label{delta0}
    \begin{aligned}
    &\check\delta A_{\alpha\alphad} = 8\Psib^-_{\alphad}\epsilon^{-\beta}
    \phib\,c_{\alpha\beta}\bpp\,,\\[3mm]
    &\check\delta \phi = 8\left[ 2\Psi^-_{\alpha}\epsilon^-_{\beta}\phib
    -A_{\alpha\alphad}\Psibm^{\alphad}\epsilon^-_{\beta} \right]
    c^{\alpha\beta}\bpp\,,\\[3mm]
    &\check\delta \Psim^\alpha =8\epsilon^-_\beta
    \left[ (\Psibm)^2-\phib D^{--} \right]c^{\alpha\beta}\bpp
    +2\iim\phib\,\epsilon^-_\gamma\partial_{\beta\alphad}A^{\gamma\alphad}\,
    c^{\alpha\beta}\bpm\\[1mm]
    &\hskip 1cm+2\iim\left[\phib\epsilon^-_\gamma
    \partial_{\beta\alphad}A^{\alpha\alphad}+\partial_{\beta\alphad}\phib
    \epsilonm^\alpha A_\gamma^\alphad\right]
    c^{\gamma\beta}\bpm\,,\\[3mm]
    &\check\delta \Psib^-_\alphad =
    4\iim\,\phib\partial_{\alpha\alphad}\phib\,\epsilon^-_\beta\,
    c^{\alpha\beta}\bpm\,,\\[3mm]
    &\check\delta  D^{--} = -8\iim\,\left[ \partial_{\alpha\alphad}
    \phib\,\epsilon^-_\beta\Psibm^{\alphad}+ \phib
    \epsilon^-_{\alpha}\partial_{\beta\alphad} \Psibm^{\alphad}
    \right]c^{\alpha\beta}\bpm\,,
    \end{aligned}
\end{equation}
where $\epsilon^{-}_\alpha = \epsilon^i_\alpha u^-_i$ and
$\epsilon^i_\alpha$ is the Grassmann $\Nc=(1,0)$ transformation
parameter. We observe here the same phenomenon as in the case of
deformed gauge transformations in Sect. 3: these variations violate the
WZ gauge due to the appearance of harmonic variables in the r.h.s., so
one is led to properly modify $\L\,$.  Moreover, in $\check\delta
V^{++}_{WZ}$ there appear additional terms linear in the Grassmann
variables
\begin{equation}
    \check\delta V^{++}_{WZ}=8 \theta^+_\alpha \epsilon^{-}_\beta (\phib)^2\,
    c^{\alpha\beta}\bpp
    -8 \bar\theta^+_{\dot\rho} A^{\dot\rho}_\alpha \epsilon^-_\beta
    \phib\, c^{\alpha\beta}\bpp +\cdots\,.
\end{equation}
Thus, in contrast to the case of gauge transformations, the WZ gauge
form of $\V$ proves to be broken by $\Nc=(1,0)$ supersymmetry
transformations with $\Lambda_c = \L$ not only in the harmonic sector
but also in the Grassmann sector.

To solve this problem, we should promote the gauge parameter $\L$ to the
one providing correct transformations laws for the components fields in
the WZ gauge.  Thus we define
\begin{equation}
\Lambda_c = \L + F_\epsilon
\end{equation}
and rewrite \eqref{unbrokensusy} in the following way
\begin{align}\label{unbrokensusy2}
\delta\V= \,&\left( \epsilon^{+\alpha}\partial_{+\alpha}
+\epsilon^{-\alpha}\partial_{-\alpha}\right)
\V-\Dpp(\L+F_\epsilon)
-\left[ \V\, ,\,(\L+F_\epsilon) \right]_\star \nonumber\\[3mm]
= \,&\delta_0\V+\check\delta\V+\hat{\delta}\V\,,
\end{align}
where
\begin{equation*}
\hat{\delta}\V= -\Dpp(F_\epsilon)-\left[ \V\, ,\,F_\epsilon
\right]_\star\,,
\end{equation*}
and
\begin{equation*}
F_\epsilon=F+\thetabp \bar{F}^-+(\thetabp)^2F^{--}
\end{equation*}
is the additional compensating gauge parameter intended for restoring
the WZ gauge. The minimal set of terms needed to eliminate the improper
harmonic and Grassmann dependence appearing in \eqref{unbrokensusy2} and
\eqref{delta0} amounts to the following form of $F_\epsilon$:
\begin{equation}
    \begin{aligned}
    F&=\theta^{+\alpha} \,f^-_\alpha\,,\\[2mm]
    \bar{F}^{-\alphad}&=\gb^{-\alphad}
    +2\iim \theta^-_{\alpha}\,\theta^{+\beta}
    \partial^{\alpha\alphad}\,f^-_{\beta}
    + \theta^{+\alpha}\,b^{--\alphad}_\alpha
    +(\thetap)^2\,\gb^{(-3)\alphad}\,,\\[2mm]
    F^{--}&= g^{--} -(\thetam)^2\theta^{+\alpha}\square f^-_\alpha
    +\iim\theta^{-\alpha}\partial_{\alpha\alphad}\gb^{-\alphad}
    +\iim \thetap^\alpha \thetam^\beta\partial_\beta^\alphad
    \bmm_{\alpha\alphad}
    +\theta^{+\alpha}f^{(-3)}_\alpha\\[2mm]&
    +\iim(\thetap)^2\theta^{-\alpha}
    \partial_{\alpha\alphad}\gb^{(-3)\alphad}
    +(\thetap)^2\,X^{(-4)}\,.
\end{aligned}
\end{equation}
The requirement that the terms in $\check\delta\V+\hat\delta\V$ which
are linear in $\theta$ and $\thetab$ must vanish in order to restore the
WZ gauge in the Grassmann sector gives rise to the following equations
\begin{subequations}
\begin{align}
\ppp f^{-\alpha} + 4  \phib f^{-}_\beta\, c^{\alpha\beta}\bpp+
8\epsilon^-_\beta (\phib)^2\, c^{\alpha\beta}\bpp=\, &0\,,\\[3mm]
\ppp \bar{g}^{-}_{\alphad}
+4\, A_{\alpha\alphad}f ^{-}_\beta\, c^{\alpha\beta}\bpp
+8\epsilon^-_\beta A_{\alpha\alphad}\phib\, c^{\alpha\beta}\bpp= \,&0\,.
\end{align}
\end{subequations}
It is straightforward to check that the proper solution to these
equations is given by
\begin{align}
    \nonumber
    f^{-\alpha}=&2\phib\,\epsilon^-_\beta\left[
    \ve^{\alpha\beta}\left( \frac{\sinh X}X \cos Z -1\right)
    -c^{\alpha\beta}\sqrt{\frac2{c^2}}\frac{\sinh X}X \sin Z
    \right]\\\nonumber
    &-4\phib^2\bmm\epsilon^+_\beta\Biggl\{
    \ve^{\alpha\beta}
    \sqrt{\frac{c^2}2} \frac{\iim}X\left[ \frac{\sinh(X+\iim Z)}{X+\iim Z}
    - \frac{\sinh (X-\iim Z)}{X-\iim Z}\right]\\\nonumber
    &-c^{\alpha\beta}
    \frac{1}{X}\left(\frac{1}{X-\iim Z}\left[\cosh(X-\iim Z) -1\right]
    +\frac{1}{X+\iim Z}\left[\cosh (X+\iim Z) -1\right]\right)\Biggr\}\,,
\end{align}
and
\begin{equation}
\bar{g}^{-}_{\alphad}=A_{\alpha\alphad}\phib^{-1}f^{-\alpha}\,.
\end{equation}
Note that these solutions are regular in $\bar\phi, c^{\alpha\beta}$ and
$b^{ik}$ as they should.

Using these expressions and requiring
\begin{equation}
\ppp\delta\phib=0\,,\;
\left(\ppp\right)^2 \delta\Psib^{-}_{\alphad}=0\,,\;
\ppp\delta A_{\alpha\alphad}=0\,, \; (\ppp)^2\delta \Psi^{-}_\alpha = 0\,,
\; (\ppp)^3\delta D^{--} = 0\,,
\end{equation}
we can explicitly find other components of $F_\epsilon$ and restore the
correct $\Nc=(1,0)$ supersymmetry transformations preserving WZ gauge.
We are planning to give the full set of these transformations in
\cite{In prep}. Here we quote only simplest ones
\begin{align}
    \delta\phib&=0\,,\\
    \delta\Psib^i_\alphad&=\left[
    \frac{2\iim}{\sqrt{b^2c^2}}\cosh X\,\sinh X\,
    c^{\alpha\beta}b^{ij}
    -\iim \cosh^2 X\,\ve^{\alpha\beta}\ve^{ij}\right]
    \epsilon_{j\beta}\partial_{\alpha\alphad}\phib\,,\\[3mm]
\delta A_{\alpha\alphad}&=
8\phib\epsilon^{i\beta}\Psib^j_{\alphad}\,b_{ij}c_{\alpha\beta}
+2\epsilon^{k}_{\alpha}\Psib_{k\alphad}\,
X\coth{X}\,.
\end{align}
These variations form the algebra which is closed modulo a gauge
transformation with the composite parameter $a_c =
-2\iim(\epsilon\cdot\eta)\bar\phi\,$:
\begin{equation*}
\left[\delta_\epsilon,\delta_\eta \right]\phib=0\,,\qquad
\left[\delta_\epsilon,\delta_\eta \right]\Psib^j_{\alphad}=0\,,
\end{equation*}
\begin{equation*}
\left[\delta_\epsilon,\delta_\eta \right]A_{\alpha\alphad}=
-2\iim(\epsilon\cdot\eta)
\left(X\coth X \right)\partial_{\alpha\alphad}\phib\,.
\end{equation*}

The $\Nc=(1,0)$ transformations are radically simplified for
the degenerate choice \eqref{3/2} with $b^2 = 0$:
\begin{equation}
\delta\Psib^-_\alphad = -i\left(\ve^{\alpha\beta} \ve^{ij}
- 4\phib c^{\alpha\beta} b^{ij}\right)
\epsilon_{j\beta}\partial_{\alpha\dot\alpha}\phib\,,
\quad \delta A_{\alpha\alphad} = 2\epsilon^{k}_\alpha\Psib_{k\alphad}
+ 8\epsilon^{i\beta}\phib \Psib^j_{\alphad}c_{\alpha\beta}b_{ij}\,.
\end{equation}

\section{Concluding remarks}
In this paper we analyzed the model of a non-singlet Q-deformed $\Nc=(1,1)$
supersymmetric U(1) gauge multiplet in harmonic superspace.
We presented exact expressions for the gauge transformation of the
fields, calculated the bosonic sector of the component action and gave a
few examples of unbroken $\Nc=(1,0)$ supersymmetry transformations in WZ
gauge. All these results have been obtained for the special decomposition
\eqref{hatcdecomp} of the SU(2)$_\text{L}\times$SU(2) deformation matrix,
namely $\hat{C}^{ij}_{\alpha\beta}=b^{ij}c_{\alpha\beta}$.
This choice contains only six degrees of freedom, in contrast to the nine
parameters of the generic non-singlet matrix
(or two versus three after choosing an appropriate frame
with respect to the broken SU(2)$_{\text{L}}\times$SU(2) symmetry \footnote{Using broken scale
$O(1,1)$ automorphism symmetry, one can further fix one parameter in both cases.}).
Let us summarize the key features of the ansatz \eqref{hatcdecomp}.
\begin{itemize}
\item It provides a unique possibility to obtain all quantities in a closed
compact form, in contrast to the generic non-singlet deformation case
\cite{Araki:2004mq, Araki:2004cv, Araki:2005pc}.
\item It realizes the maximally symmetric non-singlet deformation, with the
``Lorentz'' U(1)$_{\text{L}}$ and R-symmetry U(1) subgroups left unbroken. The
generic non-singlet deformation fully breaks SU(2)$_{\text{L}}$ and R~symmetry.
\item With the choice $b^2 =0$ it includes the important degenerate case
which preserves 3/2 of the original ${\cal N}=(1,1)$ supersymmetry
(assuming a pseudoconjugation for the deformation parameters)
\cite{Ivanov:2003te, Araki:2005pc}.
\item It directly yields the well-known Seiberg ansatz after performing
the reduction to ${\cal N}=(1/2, 1/2)$ superspace, with $c_{\alpha\beta}$
becoming the reduced deformation matrix \footnote{Let us choose e.g.
$\theta^1_\alpha$ as the left Grassmann co-ordinate of some
${\cal N}=(1/2, 1/2)$ subspace of the ${\cal N} = (1,1)$ superspace,
i.e. $\theta_1^\alpha \equiv \theta^\alpha$, assume the pseudoconjugation for
all involved quantities as in \cite{Ivanov:2003te}, and fix the relevant broken automorphism
$U(1)$ and $O(1,1)$ symmetries
of the ${\cal N} = (1,1)$ superalgebra in such a way that $b_{ik} \equiv
(b_{11}, b_{22}, b_{12}) = (1, b_{22}, 0)\,$. Then the deformation operator \eqref{YQ}
for the choice \eqref{Ccb} and $I=0$ is reduced to
$P = - \overleftarrow{\partial_\alpha} c^{\alpha\beta} \overrightarrow{\partial_\beta}
- b_{22}\overleftarrow{\partial^2_\alpha} c^{\alpha\beta} \overrightarrow{\partial^2_\beta}\,$, i.e. it is
expressed as a sum of the mutually commuting chiral Poisson operators on two
different ${\cal N}=(1/2, 1/2)$ subspaces of the ${\cal N} = (1,1)$ superspace,
with $b_{22}$ being the ``ratio'' of two Seiberg deformation matrices.
When $b_{22} = 0$, we face the case $b^2 = 0$ of Sect. 4.2, with only one ${\cal N}=(0, 1/2)$
supersymmetry broken. For $b_{22} \neq 0$, both ${\cal N}=(0, 1/2)$ supersymmetries
are broken. The parameter $b_{22}$ measures the breakdown of the second
${\cal N}=(0, 1/2)$ supersymmetry which is implicit in the ${\cal N}=(1/2, 1/2)$
superfield formulation based on the  superspace $(x^m, \theta^\alpha, \bar\theta^{\dot\alpha})\,$.
Recall that within the standard complex conjugation the reduction to ${\cal N}=(1/2, 1/2)$ superspace
makes no sense since the latter is not closed under such conjugation \cite{Ivanov:2003te}.}.
\end{itemize}

The bosonic action has a structure similar to the singlet
Q-deformed one calculated in \cite{Araki:2004de,Ferrara:2004zv}, in the sense that
the gauge field develops a non-trivial interaction with one of the
$\Nc=(1,1)$ vector multiplet scalar field, i. e. $\phib$.  This
interaction is the exact zero-fermion limit of the full component
interaction lagrangian which, as follows from the superfield setup we started with,
by construction breaks some fraction of the original supersymmetry. In the general
case of $c^2 \neq 0, b^2 \neq 0$ the full component action  should break the original $\Nc=(1,1)$
supersymmetry by half, i.e. down to $\Nc=(1,0)\,$, whereas the
degenerate choice $b^2=0$ preserves  the fraction $3/2$ of $\Nc=(1,1)$
supersymmetry, as noticed in \cite{Ivanov:2003te} and discussed in more
detail in a recent paper \cite{Araki:2005pc}.  While the characteristic
object of the Q-deformed $\Nc=(1,1)$ gauge theory with the singlet
deformation matrix is the polynomial factor $(1 + 4I\bar\phi)$
\cite{Araki:2004de,Ferrara:2004zv}, where $I$ is the deformation parameter, in the
considered case there naturally appear {\it hyperbolic\/} functions of the
argument $X=2\phib\sqrt{b^2c^2}\,$.  Though the intrinsic reason of this
mysterious appearance  of the hyperbolic functions in the case of
QNS-deformation with the product deformation matrix is so far unclear,
it hopefully could be understood after clarifying possible relation of
this sort of nilpotent deformation to some non-trivial backgrounds in
string theory. Anyway, it is the manifestly supersymmetric harmonic
superfield approach which allowed us to reveal these non-trivial
structures at the component level: it would hardly be possible to
exhibit them using from the very beginning the component approach or any
approach based on the ordinary superfields.

More details of this special QNS-deformation of $\Nc=(1,1)$ gauge
theory, in particular, the total component action with fermions, will be
given in our forthcoming paper \cite{In prep}. As for the possible
further developments, it would be interesting to study implications of
this deformation in non-abelian $\Nc=(1,1)$ gauge theory and in the
models including hypermultiplets, along the lines of refs.
\cite{Ferrara:2004zv,Ivanov:2004hc}. The study of quantum and geometric
properties of these models, equally as revealing their possible
phenomenological applications, e.g. as providing specific mechanism of
the soft supersymmetry breaking, surely deserve further attention.

Finally, it would be interesting to treat the case of the generic non-singlet
deformation matrix $\hat{C}^{ij}_{\alpha\beta}$ as a perturbation
around the non-trivial ansatz \eqref{hatcdecomp} rather than around
the undeformed limit. In this way one can hope to find a closed formulation
of the general QNS-deformed theory.

\section*{Acknowledgments}
This work was partially supported by the joint DAAD -- Fundaci\'on
Gran Mariscal de Ayacucho scholarship (L.Q.) and fellowship (A.D.C.).
A.D.C.~and L.Q.~want to thank Prof.~Alvaro Restuccia and Nicolas Hatcher
for important discussions.
E.I.~is grateful to Boris Zupnik for useful discussions.
E.I.~and O.L.~acknowledge a support from the DFG grants 436~RUS~113/669-2
and Le-838/7. The work of E.I.~was partially supported by the RFBR
grants 03-02-17440 and 04-0204002, the NATO grant PST.GLG.980302 and
a grant of Heisenberg-Landau program.

\appendix

\section{Notation and conventions}
\label{Appendix1}

In this Appendix we review the basics of $\Nc=(1,1)$ Euclidean harmonic
superspace which we use throughout the paper.  For a deeper treatment of
this subject we refer to \cite{Galperin1,Galperin:2001uw} (see also
\cite{Ivanov:2003te,Ferrara:2004zv,Ivanov:2004hc}).

The Euclidean harmonic superspace is defined as the product
\begin{equation}
    {\HR}^{4+2|8}=\mathbbm{R}^{4|8}\times\frac{\mbox{SU(2)}}{\mbox{U(1)}}\,,
\end{equation}
where $\mathbbm{R}^{4|8}$ is the $\Nc=(1,1)$ Euclidean
superspace and SU(2) is the R-symmetry (automorphism) group of
$\Nc=(1,1)$ superalgebra.  The topology of this superspace is
$\mathbbm{R}^{4|8}\times S^2$ and it is parametrized by the standard $(4
+8)$ coordinates $(x^{\alpha\dot\alpha}, \theta^\alpha_i,
\bar\theta^{\dot\alpha i})$ of the superspace $\mathbbm{R}^{4|8}$ and
the SU(2) harmonic variables representing sphere $S^2 \sim $ SU(2)/U(1)
and denoted by $u^{\pm}_i\,$.\footnote{Throughout the paper, Greek
indices $\alpha, \dot\alpha$ are spinorial indices of the group
$Spin(4)=$ SU(2)$_{\text{L}}\times$ SU(2)$_{\text{R}}$ and Latin indices
$i, j$ are the doublet indices of the R-symmetry group SU(2).  Both
sorts of indices are raised and lowered with the skew-symmetric metric
$\ve^{\alpha\beta}, \ve^{\dot\alpha\dot\beta}, \ve^{ik}$, e.g.
$u^{\pm}_{k}=\ve_{kj}u^{\pm\,j},\; \,u^{\pm\,k}=\ve^{kj}u^{\pm}_j$,
where $\ve_{12}=1\,,\;\ve^{\alpha\beta}\ve_{\beta\gamma}
=\delta^{\alpha}_{\gamma}\,, \quad \mbox{etc}\,$.} The harmonic
variables  are defined by the completeness relation
\begin{align}
    u^+_{i}\,u^-_j-u^+_{j}\,u^-_i=\ve_{ij}\,.
    \label{redident}
\end{align}
Their symmetrized products
\begin{equation*}
u^+_{(i_1}\ldots\,u^+_{i_n}u^-_{j_1}\ldots\,u^-_{j_m)}
\end{equation*}
form a complete basis of functions on the sphere $S^2$.

While in the {\it central basis\/} the harmonic superspace is represented
by the coordinates $(x^{\alpha\alphad},\theta^{\alpha}_i,$
$\thetab^{\alphad i},u^\pm_i)$, its {\it analytic basis\/} is defined as
the coordinate set $(x^{\alpha\alphad}_A,\theta^{\alpha\pm},$
$\thetab^{\alphad\pm}, u^{\pm}_i)\,,$ where
\begin{align}
x^{\alpha\alphad}_A=&x^{\alpha\alphad}-4\iim\theta^{\alpha\,i}
\thetab^{\alphad\,j}u^-_{(i}\,u^+_{j)}\, ,\cr
\theta^{\alpha\pm}=
&\theta^{\alpha\,k}u^{\pm}_k\,,\quad\thetab^{\alphad\pm}=
\thetab^{\alphad\,k}u^{\pm}_k\,.
\end{align}

The covariant spinor derivatives in the analytic basis are defined as
\begin{align}
D^k_\alpha\,u^+_k=&D^+_\alpha=\partial_{-\alpha}\,,
\qquad D^k_\alpha\,u^-_k=D^-_\alpha=-\partial_{+\alpha}+
2\iim{\thetab}^{-\alphad}\partial_{\alpha\alphad}\,,\cr
\bar{D}^k_{\alphad}\,u^+_k=
&\bar{D}^+_\alphad=\bar{\partial}_{-\alphad}\,,
\qquad \bar{D}^k_\alphad\,u^-_k=
-\bar{D}^-_{\alphad}=-\bar{\partial}_{+\alphad}-
2\iim\theta^{-\alpha}\partial_{\alpha\alphad}\,,
\end{align}
where
\begin{equation*}
\partial_{\pm\alpha}=
\frac{\partial}{\partial\,\theta^{\pm\alpha}}\,,
\quad
\bar{\partial}_{\pm\alphad}=
\frac{\partial}{\partial\,\thetab^{\pm\alphad}}\,,
\quad\partial_{\alpha\alphad} = 2 \frac{\partial}{\partial x^{\alpha\alphad}}
\end{equation*}
and $D^k_\alpha\,, \bar D^k_{\alphad}$ are spinor derivatives in the
central basis (their explicit form is not needed for us).

An important ingredient of the harmonic superspace is the covariant
derivatives with respect to the harmonic variables. In the analytic
basis they are
\begin{eqnarray}
\Do_A &=&\partial^0 + \theta^{+\alpha}\partial_{+\alpha}
-\theta^{-\alpha}\partial_{-\alpha}+
{\thetab}^{+\alphad}\bar{\partial}_{+\alphad}-
{\thetab}^{-\alphad}\bar{\partial}_{-\alphad}\,,\nn
\Dpp_A &=& \ppp
-2\iim\theta^{+\alpha}{\thetab}^{+\alphad}\partial_{\alpha\alphad}
+\theta^{+\alpha}\partial_{-\alpha}
+{\thetab}^{+\alphad}\bar{\partial}_{-\alphad}\,,\nn
\Dmm_A &=& \pmm
-2\iim\theta^{-\alpha}{\thetab}^{-\alphad}\partial_{\alpha\alphad}
+\theta^{-\alpha}\partial_{+\alpha}
+{\thetab}^{-\alphad}\bar{\partial}_{+\alphad}
\end{eqnarray}
with
\begin{equation*}
\partial^0=u^{+i}\frac{\partial}{\partial\,u^{+i}}
-u^{-i}\frac{\partial}{\partial\,u^{-i}}\,,
\quad\text{and}\quad
\partial^{\pm\pm}=
u^{\pm i}\frac{\partial}{\partial\,u^{\mp i}}\, .
\end{equation*}
They form an SU(2) algebra:
\begin{equation*}
\left[\,\Dpp\,,\,\Dmm\,\right]=\,\Do\,,\quad
\left[\,\Do\,,\,\Dpmpm\,\right]=\pm\,2\,\Dpmpm\,.
\end{equation*}

The \emph{left-chiral basis} of the harmonic superspace is represented
by the coordinates $(x^{\alpha\alphad}_L,\,\theta^{\pm\alpha},$ $
\thetab^{\pm\alphad}, u^\pm_i )\,,$ where
\begin{align}
x^{\alpha\alphad}_L=&
x^{\alpha\alphad}_A+4\iim\theta^{-\alpha}\thetab^{+\alphad}\,.
\end{align}
In this basis, the  differential operators used throughout the paper
are written as
\begin{eqnarray}
&&D^+_\alpha= \partial_{-\alpha}
+2\iim\thetab^{+\alphad}\partial_{\alpha\alphad}\,, \quad
D^-_\alpha= -\partial_{+\alpha}
+2\iim\thetab^{-\alphad}\partial_{\alpha\alphad}\,,\quad
\bar{D}^+_\alphad= \bar{\partial}_{-\alphad}\,, \quad
\bar{D}^-_\alphad= -\bar{\partial}_{+\alphad}\,, \nn
&& \Dpp= \partial^{++} + \theta^{+\alpha}\partial_{-\alpha}+
\thetab^{+\alphad}\bar{\partial}_{-\alphad}\,,\quad
\Dmm=\partial^{--} + \theta^{-\alpha}\partial_{+\alpha}+
\thetab^{-\alphad}\bar{\partial}_{+\alphad}\,, \nn
&& Q^+_\alpha= \partial_{-\alpha}\,, \quad
Q^-_\alpha= -\partial_{+\alpha}\,.
\end{eqnarray}

\section{Some useful harmonic integrals}
\label{harmints}

In this Appendix we give explicit formulas for some harmonic integrals
appearing in our calculations.

The integration over harmonics is fully specified by the two rules
\cite{Galperin:2001uw}
\begin{equation*}
(a)\,\int du \, 1 = 1\,, \qquad
(b)\,\int du\, u^+_{(i_1}\ldots\,
u^+_{i_n}u^-_{j_1}\ldots\,u^-_{j_m)} = 0\,.
\end{equation*}
This means, in particular, that the harmonic integral of any object of
the form $\partial^{++}f^{--}$ or $\partial^{--}f^{++}$ equals to zero,
i.e. one can integrate by parts.

Using this property and eqs. \eqref{evenoddb}, it is easy to compute
\begin{align}
\int\,du\, (\bpm)^{2(n+1)} =
\frac{(-1)^{n+1}}{2n+3}\left(\frac{b^2}2\right)^{n+1},\quad\quad
\int\,du\,(\bpm)^{2n+1}= 0\,,
\end{align}
whence, recalling that $Z=2\phib\sqrt{2\,c^2}\,\bpm$ and
$X=2\phib\sqrt{b^2\,c^2}\,$,
\begin{eqnarray}
&& \int du\,\cos Z = \frac{\sinh X}{X}\,,  \label{I1} \\
&& \int du \,Z\sin Z = \frac{\sinh X - X\cosh X}{X}\, , \label{I2} \\
&& \int du\, Z^2\,\cos Z =
    \frac{2X\cosh X - 2\,\sinh X - X^2\,\sinh X}{X}\, , \label{I3} \\
&& \int du\,Z^3\,\sin Z =
    \frac{X^3\cosh X - 3\,X^2 \sinh X + 6\,X\,\cosh X - 6\,\sinh X}{X}\,,
    \label{I4}\\
&& \int du\, Z^4\,\cos Z=
-4(6+X^2)\cosh{X}+\frac{(24+12 X^2+X^4)\sinh{X}}{X}\,, \;\;\mbox{etc}\,.
\label{I5}
\end{eqnarray}
The easiest way to compute integrals \eqref{I2} - \eqref{I5} is to
introduce a real parameter $\alpha$ into \eqref{I1} as $Z \rightarrow
\alpha Z$, $X \rightarrow \alpha X$ and to repeatedly differentiate both
sides of \eqref{I1} with respect to $\alpha\,$.

Another type of harmonic integrals include some object $A^{+-} =
A^{(ik)}u^+_iu^-_k$:
\begin{eqnarray}
&& \int\,du\, A^{+-}\;(\bpm)^{2n+1}=
\frac{(-1)^{n+1}}{2(2n+3)}\,(A\cdot\,b)\,
\left( \frac{b^2}{2}\right)^n\,,\label{I6} \\
&&\int\,du\,(A^{+-})^2\;(\bpm)^{2n}= (-1)^{n+1}
\frac{\left[A^2\;b^2+2n\;(A\cdot b)^2 \right]}{4\;(2n+1)\;(2n+3)}
\left( \frac{b^2}{2}\right)^{n-1}\,, \label{I7} \\
&& \int\,du\, A^{+-}\;(\bpm)^{2n}=0\,,\qquad
\int\,du\,(A^{+-})^2\;(\bpm)^{2n+1}=0\,. \label{I8}
\end{eqnarray}
The vanishing of integrals in \eqref{I8} directly follows from the fact
that they should be SU(2) invariants and the observation that it is
impossible to form SU(2) invariants from the given sets of traceless
tensors $A^{(ik)}$ and $b^{(ik)}$. Integrals \eqref{I6} and \eqref{I7}
can be directly computed using the identities of the type \eqref{229}
following from the completeness relation \eqref{redident}.


\end{document}